\documentstyle[preprint,aps,fixes,epsf,eqsecnum]{revtex}
\tightenlines

\makeatletter           % Switch to floating figures
\@floatstrue
\def\figure{\let\@capwidth\columnwidth\@float{figure}}
\let\endfigure\end@float
\@namedef{figure*}{\let\@capwidth\textwidth\@dblfloat{figure}}
\@namedef{endfigure*}{\end@dblfloat}
\def\table{\let\@capwidth\columnwidth\@float{table}}
\let\endtable\end@float
\@namedef{table*}{\let\@capwidth\textwidth\@dblfloat{table}}
\@namedef{endtable*}{\end@dblfloat}

\makeatother

\def\includegraphics[scale=#1]#2{\centerline{\def\epsfsize ##1##2{#1##1}\epsfbox{#2}}}

\begin{document}

\preprint{UW/PT 02--06}

\title{Photon and Gluon Emission in Relativistic Plasmas}

\author{Peter Arnold}
\address
    {%
    Department of Physics,
    University of Virginia,
    Charlottesville, Virginia 22901
    }%
\author{Guy D. Moore and Laurence G. Yaffe}
\address
    {%
    Department of Physics,
    University of Washington,
    Seattle, Washington 98195
    }%

\date{\today}

%\pacs{Valid PACS appear here}

%\keywords{Suggested keywords}%Use showkeys class option if keyword
                              %display desired

\maketitle

\begin{abstract}
    We recently derived, using diagrammatic methods,
    the leading-order hard photon emission rate in ultra-relativistic plasmas.
    This requires a correct treatment of multiple scattering effects which
    limit the coherence length of emitted radiation
    (the Landau-Pomeranchuk-Migdal effect).
    In this paper, we provide a more physical derivation of this result,
    and extend the treatment to the case of gluon radiation.
\end{abstract}

%%%%%%%%%%%%%%%%%%%%%%%%%%%%%%%%%%%%%%%%%%%%%%%%%%%%%%%%%%%%%%%%%%%%%%%%%%%%%%%

\def\Re{{\rm Re}}
\def\Im{{\rm Im}}
\def\PiL{\Pi_{\rm L}}
\def\PiT{\Pi_{\rm T}}
\def\rhoT{\rho_{\rm T}}
\def\rhoL{\rho_{\rm L}}
\def\PL{P_{\rm L}}
\def\PT{P_{\rm T}}
\def\ret{{\rm Ret}}
\def\adv{{\rm Adv}}
\def\sgn{{\rm sign}}
\def\tpp{\tilde {\bf p}_\perp}
\def\tqp{\tilde {\bf q}_\perp}
\def\tpsq{{\tilde{p}_{\perp}^2}}
\def\tqsq{{\tilde{q}_{\perp}^2}}
\def\tbf{\tilde{\bf f}}
\def\dEtw{{\widetilde{\delta E}\rule{0ex}{2.3ex}}}
\def\dEhat{\widehat{\delta E}}
\def\ppar{{p_\parallel}}
\def\qpar{{q_\parallel}}
\def\ipa{inelastic pair annihilation}
\def\twotwo{2 \leftrightarrow 2}

\def\pxk{{\bf h}}
\def\j{{\bf j}}
\def\x{{\bf x}}
\def\y{{\bf y}}
\def\p{{\bf p}}
\def\q{{\bf q}}
\def\k{{\bf k}}
\def\v{{\bf v}}
\def\u{{\bf u}}
\def\l{{\bf l}}
\def\E{{\bf E}}
\def\B{{\bf B}}
\def\D{{\bf h}}
\def\h{{h}}
\def\magp{|{\bf p}|}
\def\f{{\overline{f}}}
\def\C{{\cal C}}

\def\A{{\rm A}}
\def\B{{\rm B}}
\def\V{{\rm V}}
\def\T{{\rm T}}
\def\LHS{{\rm LHS}}
\def\naBla{{\bf \nabla}}
\def\ij{{i \cdots j}}

\def\ca{C_{\rm A}}
\def\cf{C_{\rm F}}
\def\da{d_{\rm A}}
\def\ta{T_{\rm A}}
\def\tf{T_{\rm F}}
\def\df{d_{\rm F}}
\def\nf{N_{\rm f}}
\def\nc{N_{\rm c}}
\def\mD{m_{\rm D}}
\def\nb{n_{\rm B}}
\def\nq{n_{\rm q}}
\def\trans{\top}

\def\tfrac#1#2{{\textstyle {#1 \over #2}}}
\def\dfrac#1#2{{\displaystyle {#1 \over #2}}}

\def\sigT{\sigma_\T}
\def\grad{\mbox{\boldmath$\nabla$}}
\def\half{{\tfrac{1}{2}}}
\def\CA{C_{\rm A}}
\def\eps{\epsilon}
\def\gammaE{\gamma_{\rm E}}
\def\Tan{{\rm Tan}}
\def\MSbar{$\overline{\hbox{MS}}$}
\def\gs{g_{\rm s}}
\def\gw{g_{\rm w}}
\def\alphaw{\alpha_{\rm w}}
\def\alphas{\alpha_{\rm s}}
\def\alphaEM{\alpha_{\scriptscriptstyle \rm EM}}
\def\bPhi{\mbox{\boldmath$\Phi$}}
\def\c{{\bf c}}
\def\blambda{\mbox{\boldmath$\lambda$}}
\def\FAC{{\hbox{\tiny FAC}}}
\def\NLLO{{\hbox{\tiny NLLO}}}
\def\Const{{\cal C}}
\def\chie{\chi^{e^{+}\!}}

\def\gsim{\mbox{~{\raisebox{0.4ex}{$>$}}\hspace{-1.1em}
	{\raisebox{-0.6ex}{$\sim$}}~}}
\def\lsim{\mbox{~{\raisebox{0.4ex}{$<$}}\hspace{-1.1em}
	{\raisebox{-0.6ex}{$\sim$}}~}}
%%% Avoids tiny looking <,> in the <~ and >~ symbols. %%%
\def\drangle{\rangle\!\rangle}
\def\dlangle{\langle\!\langle}
\def\bigdrangle{\bigr\rangle\!\bigr\rangle}
\def\bigdlangle{\bigl\langle\!\bigl\langle}
\def\deltaS{\delta^{S_2}}
\hyphenation{brems-strah-lung}

\def\deltaC{\delta\hat C}
\def\diffC{A}
\def\ffh{{\rm f\bar fh}}
\def\ffhc{{\rm f\bar fhc}}
\def\fh{{\rm fh}}

%% new definitions needed for this new section

\def\Biggdlangle{\Bigg\langle\!\!\!\Bigg\langle}
\def\Biggdrangle{\Bigg\rangle\!\!\!\Bigg\rangle}
\def\biggdlangle{\bigg\langle\!\!\!\bigg\langle}
\def\biggdrangle{\bigg\rangle\!\!\!\bigg\rangle}
\def\Bigdlangle{\Big\langle\!\!\Big\langle}
\def\Bigdrangle{\Big\rangle\!\!\Big\rangle}
\def\brem{bremsstrahlung}
\def\vac{{\rm vac}}
\def\fff{{\cal S}}
\def\ff{S}
\def\blparen{\mbox{\boldmath$($}}
\def\brparen{\mbox{\boldmath$)$}}
\def\LPM{{\scriptscriptstyle \rm LPM}}

\def\kpar{{k_\parallel}}
\def\epar{{{\bf e}_\parallel}}
\def\bDelta{\mbox{\boldmath$\Delta$}}

%%%%%%%%%%%%%%%%%%%%%%%%%%%%%%%%%%%%%%%%%%%%%%%%%%%%%%%%%%%%%%%%%%%%%%%%%%%%%%%

\section {Introduction}

The rate of photon emission in a high-temperature QCD plasma
is a problem of some theoretical interest
\cite{photon1,photon2,photon3,photon4,photon5,photon6,photon7,photon8,%
Kapusta,Baier,Gelis1,Gelis2,Gelis3,SteffenThoma},
due in part to the hope that hard photon emission
will be a useful diagnostic probe of heavy ion collisions
\cite{heavyion1,heavyion2}.
Understanding the analogous rate of gluon emission in a hot QCD plasma
is required for computing thermalization and transport processes
\cite {paper3,Zakharov,BDMPS,LPM_QCD1,LPM_QCD2,LPM_QCD3,LPM_QCD4,gyulassy&wang}.%

\begin{figure}[t]
  \vspace*{-0.25em}
  \includegraphics[scale=0.38]{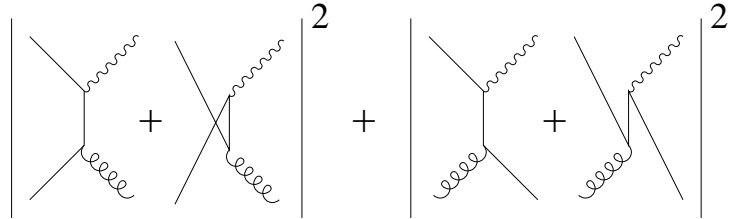}
  \medskip
  \caption{
    Two-to-two particle processes contributing to the leading order
    photon emission rate.
    Time may be viewed as running from left to right.
    \label{fig:2to2}
  }
\end{figure}

\begin{figure}
  \vspace*{-0.25em}
  \includegraphics[scale=0.38]{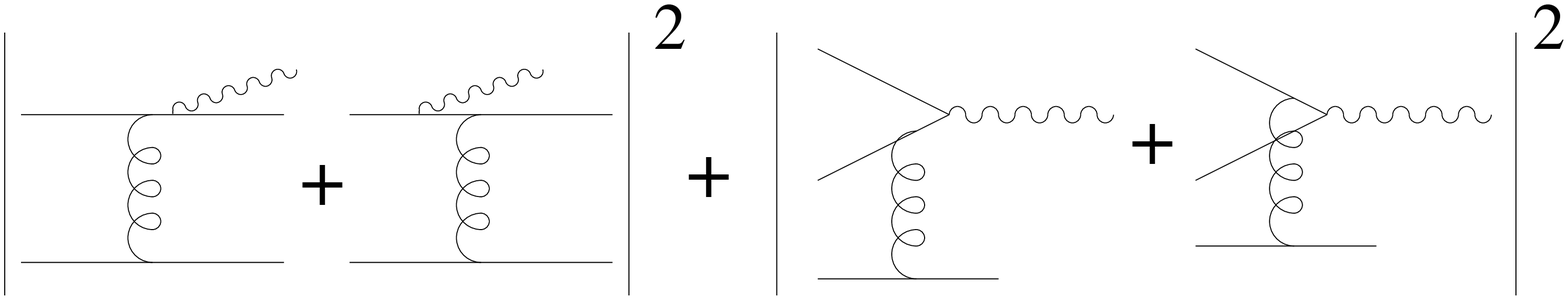}
  \medskip
  \caption{
    Bremsstrahlung and pair production contributions to photon emission.
    The bottom line in each diagram can represent either a quark or a gluon.
    %Time may again be viewed as running from left to right.
    \label{fig:inelastic}
  }
\end{figure}

\begin{figure}
  \includegraphics[scale=0.40]{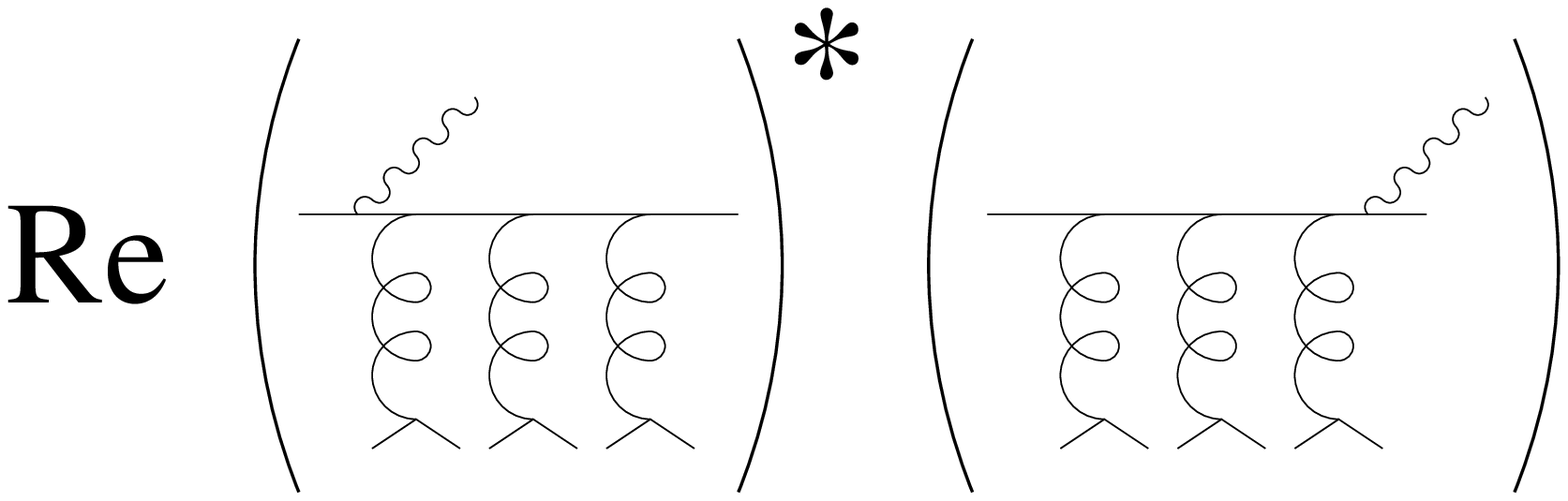}
  \caption{
    An interference term, involving amplitudes for photon emission before and
    after multiple scattering events,
    which contributes to the leading order emission rate.
    \label{fig:intromultiple}
  }
\end{figure}

We will make the simplifying assumption that the temperature $T$ is so
large that the running strong coupling constant $\gs(T)$ can be
treated as small, and focus on the leading-order (in $\alphas$)
behavior of the photon or gluon emission rate.%
\footnote
    {%
    In other words, we will neglect all sub-leading corrections to
    the emission rate suppressed by additional powers of $\alphas$,
    as well as power corrections suppressed by $\Lambda_{\rm QCD}/T$
    or $m_q/T$.
    }
Evaluating just the leading weak coupling behavior,
even in this asymptotically high temperature regime,
is non-trivial.
For the photon emission rate,
one might expect that it would be sufficient to
compute just the diagrams shown in Fig.~\ref{fig:2to2},
which describe lowest order two-to-two particle scattering processes.
But it turns out that naive perturbation theory
(augmented by the standard resummation of hard thermal loops)
does not suffice to calculate the leading-order hard photon emission rate.
Fig.\ \ref{fig:inelastic} shows 
\brem\ and pair production processes which also contribute
to the on-shell photon emission rate at leading order.
The essential problem is that the internal time scale associated with
these processes is comparable to the mean free time for soft scattering
with other particles in the plasma.
Consequently, even at leading order, the photon emission rate is
sensitive to processes involving multiple scatterings occurring during
the emission process.
This is known as the Landau-Pomeranchuk-Migdal (LPM) effect.
Diagrammatically, it manifests as the presence of interference terms
involving multiple collisions, such as depicted in
Fig.\ \ref{fig:intromultiple},
which parametrically are equally as important as those of
Fig.\ \ref{fig:inelastic}.
In the nearly-collinear limit,
the extra explicit factors of $\gs$ turn out to be
canceled by a combination of parametrically
large enhancements from the internal quark propagators
and soft exchanged gluons.

In Ref.\ \cite{paper1},
we showed how to account for the LPM effect in photon emission by identifying and
summing the appropriate infinite class of diagrams which contribute
to the leading order emission rate,
and in Ref.\ \cite{paper2} we solved the resulting integral equations numerically.
Our analysis in Ref.\ \cite {paper1} involved
quite detailed power counting of diagrams and relied upon some rather
unintuitive relationships found by Wang and Heinz \cite{HW1}
between real-time thermal 4-point Green's functions
in the r/a (Keldysh) formalism.
One goal of the present article is to
reproduce our previous results in a manner that more clearly highlights
the essential physics of the result.
We will embrace physical arguments wherever possible,
leaving to our previous work \cite{paper1} the
more technical justification of our results.
Our second goal will be to extend the treatment to the case
of gluon emission.

We are hardly the first people to discuss applications of the LPM effect,
either qualitatively or quantitatively
\cite {Gelis3,Zakharov,BDMPS,LPM_QCD1,LPM_QCD2,LPM_QCD3,LPM_QCD4,%
gyulassy&wang,LP,M1,M2,LPM_QED1,LPM_QED2,LPM_QED3}.%
\footnote{
  For a review of many aspects of the LPM effect, with an emphasis on
  the theory and experiment of
  high-energy collisions of electrons with atomic matter,
  see Ref.\ \cite{klein}.
}
However, discussions prior to our work \cite{paper1}
have almost always focused on systems where the scatterers
with which the emitting particle interacts are static.
(The scatterers are represented
by the bottom lines in Figs.\ \ref{fig:inelastic} and \ref{fig:intromultiple}.)
But typical scatterings in an ultra-relativistic plasma
involve interactions with excitations which
are themselves moving at nearly the speed of light.
They produce dynamically screened color electric and magnetic
fields which form a fluctuating background field in which \brem,
pair annihilation, and the LPM effect take place.

Our discussion will be largely self-contained and will not directly rely
on previous treatments of the LPM effect in other contexts.
One could, presumably, explicitly mimic previous discussions such
as the seminal 1955 analysis of Migdal \cite {M1,M2}, and suitably
generalize the treatment to non-static scattering.
However, it is rather challenging to follow in detail all of the
assumptions and approximations made in Migdal's quantum mechanical treatment.
We believe that readers interested in understanding the effect can benefit
from several different formulations.
Moreover, our discussion of scales and approximations will be tailored
to the particular problem at hand,
namely hard photon (and later gluon) emission in ultra-relativistic plasmas.

We will focus on the contributions of \brem\ and pair annihilation to the rate
of photon (or gluon) emission.
We will not explicitly discuss contributions to the emission rate
from the $\twotwo$ processes of Fig.\ \ref{fig:2to2},
which are of the same order but do not require a treatment of the LPM effect.
These $\twotwo$ contributions to the photon emission rate
are calculated in Ref.\ \cite{Kapusta,Baier,paper2}.%
\footnote{
   Only the $k \gg T$ limit is addressed
   in Refs.\ \cite{Kapusta,Baier}.
}
Diagrammatically, the \brem\ and pair
annihilation processes of Fig.~\ref{fig:inelastic}
look like they are suppressed by an
explicit factor of $\gs^2$ compared to the $\twotwo$ processes of
Fig.\ \ref{fig:2to2}.
However, the diagrams of Fig.\ \ref{fig:inelastic} have a collinear
enhancement, associated with emission of the photon.
This enhancement is produced by a would-be collinear singularity which
is cut off by the
effective thermal mass
(of order $\gs T$)
and width of the fermions,
yielding a parametrically large near on-shell enhancement which causes
these processes to be the same order (up to a logarithm of $\gs$)
as those of Fig.\ \ref{fig:2to2}.

We will use the following nomenclature in our discussion.
The particles corresponding to the lower lines in
Figs.\ \ref{fig:inelastic} or \ref{fig:intromultiple} will be
called the scatterers.  The particle which emits the photon in
\brem, or the particle/anti-particle pair which annihilate in
pair annihilation, will be called the (photon) emitters.
Each separate gluon exchange in Figs.\ \ref{fig:inelastic} and
\ref{fig:intromultiple} will be referred to as a ``scattering''
of the emitters.%
\footnote{
   When distinguishing emitters from scatterers, one need not worry
   about the fundamental indistinguishability of quarks in, say,
   quark-quark scattering, because the scattering is dominated
   by small angle collisions, and the photon is emitted nearly collinear
   with the emitter(s).  So, at leading order, emitters and scatterers
   are distinguishable simply by whether or not their direction of motion
   is nearly aligned with the emitted photon.
}

\begin{figure}
  \includegraphics[scale=0.70]{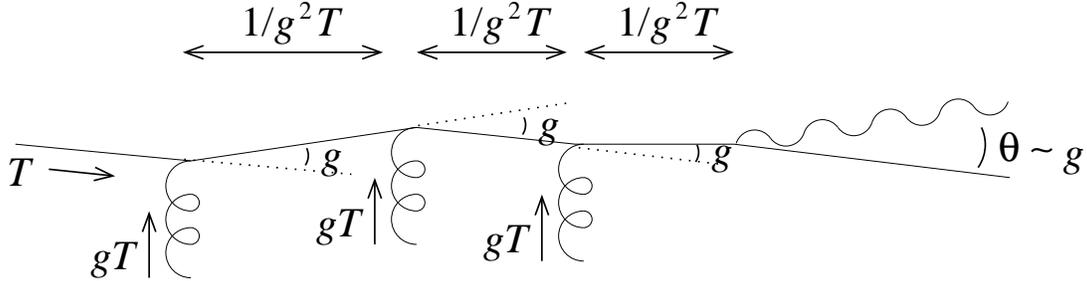}
  \caption{
    Orders of magnitude of various momentum, distance, and angular scales
    associated with \brem\ of a photon with momentum of order $T$.
    $g$ stands for the strong coupling $\gs$.
    \label{fig:scales}
  }
\end{figure}

Before we begin, we first summarize the basic scales associated
with \brem\ and pair production of hard photons via the processes of
Fig. \ref{fig:inelastic}.
For simplicity, we restrict attention to photons
with momenta parametrically of order $T$,
which is the typical momentum of particles in the ultra-relativistic plasma.
We will throughout treat electromagnetic couplings as small compared
to the QCD coupling, and we will ignore thermal effects on the
propagation of the photon (which are proportional to $\alphaEM$).
The basic kinematics, summarized pictorially
in Fig.\ \ref{fig:scales} for \brem, turn out to be as follows.

\begin {itemize}
\item
The typical momenta of the emitters and scatterers is order $T$.

\item
The typical momentum transfer $q$ of an exchanged gluon responsible for
scattering during the emission of the photon
is of order $\gs T$, which is also the order of
the inverse Debye screening length for color.
The angle of deflection in such a collision is of order
this momentum transfer divided by the emitter momentum of
order $T$, hence $\theta \sim \gs$.

\item
Processes contributing to the leading-order emission rate are dominated
by nearly collinear emission of the photon.
The corresponding internal quark lines in Fig.\ \ref{fig:inelastic} are
nearly on-shell, with energies off-shell by an amount $\delta E$ of order
$\gs^2 T$.  Fourier transforming, this implies that these
processes have time durations of order $(\gs^2 T)^{-1}$.
This is known as the {\it formation time}\/ of the photon.

\item
The formation time scale $(\gs^2 T)^{-1}$ is also the order of the mean free
time between collisions with momentum transfers of order $\gs T$.  This is
why multiple collisions cannot be treated independently and interferences
such as Fig.\ \ref{fig:intromultiple} must be included at leading order.

\item
The typical angle between the directions of the photon and the emitter(s)
(be they initial state or final state emitters) is $\theta \sim \gs$.

\item
The typical angle between the directions of the emitter and a scatterer
is $O(1)$.

\end {itemize}
A brief review of how to obtain these scales is given in Sec.\ I.B of
Ref.\ \cite{paper1}.  Here, we will take them as our starting point and
proceed to discuss how to sum up the effects of multiple collisions.

Certain physical processes in high temperature gauge theories
(such as the rate of baryon number violation in hot electroweak theory)
are sensitive to ``ultra-soft'' collisions which are mediated by
low-frequency non-Abelian magnetic fluctuations with momentum transfers
of order $\gs^2 T$.
The dynamics of such ultra-soft collisions are intrinsically non-perturbative.
One may check {\it a posteriori},
by inspection of the final answers we will produce,
that the leading-order dynamics relevant for hard photon (or gluon) production
is not sensitive to ultra-soft collisions.
Alternatively, the reader may find detailed qualitative and
diagrammatic discussions of this point in Ref.\ \cite{paper1}.

For simplicity, we will restrict attention to the case of zero chemical
potential.  The generalization of results for photon emission to non-zero
chemical potential may be found in Ref.\ \cite{paper1}.
In the next section, we will mainly focus on \brem\ and reformulate
the problem as that of \brem\ from an emitter that is propagating
through a classical random background color gauge field.
We will show how simple physical considerations of localization
of particles in both space and momentum lead to a description of the LPM
effect in \brem\ as an infinite sum of ladder diagrams.
In section \ref{sec:integral}, we will discuss how to convert the required sum
of ladder diagrams into a Boltzmann-like integral equation---a task
previously carried out in Ref.\ \cite{paper1}, but which we will
repeat using the formalism of this paper, in a way that more
clearly displays which portions of the relevant physics depend on the
nature of the emitter
({\it e.g.}\ an actual Dirac quark, a fictitious scalar quark,
or something else)
and which physics does not.
In section \ref{sec:pair}, we then analyze the closely related process
of pair production, and we combine these results for photon emission
in section \ref{sec:final}.
Finally, in section \ref {sec:gluon} we discuss the generalization to
gluon emission.

% ============================================================================

\section{Random backgrounds and ordering of interactions}

The differential photon emission rate
(per unit volume), at leading order in $\alphaEM$,
is given by the well-known relation
\begin {equation}
    d\Gamma_\gamma = \frac{d^3\k}{(2\pi)^3 \, 2|\k|} \>
    \sum_{a=1,2} \epsilon^\mu_{(a)}(\k)^* \, \epsilon^\nu_{(a)}(\k) \,
    W_{\mu\nu}(K) \,,
\label {eq:dGamma}
\end{equation}
where $K=(k^0,\k)=(|\k|,\k)$ denotes the null photon 4-momentum,
the $\epsilon$'s
represent a basis of transverse polarizations for the photon, and
$W_{\mu\nu}(K)$ is the Wightman electromagnetic current-current
correlator, 
\begin {equation}
    W_{\mu\nu}(K) = \int d^4x \> e^{-i K x} \, 
    \left\langle j_\mu(0) j_\nu(x) \right\rangle .
\label {eq:Wmunu}
\end {equation}
%% LGY:  This form is equivalent to $K -> -K$ redefinition of previous defn.
%% This form makes the connection with $S_\nu(\p;\k)$ as transparent as
%% possible (at least to me).
We use a metric with signature $({-}{+}{+}{+})$.
Because the correlator is contracted with photon polarization
vectors in (\ref{eq:dGamma}), one need only consider
the case where $\mu$ and $\nu$ are spatial indices in what follows.
As always, $\langle \cdots \rangle$ denotes an expectation value
in whatever density matrix ${\cal P}$ is of interest,
which in our case is a thermal ensemble
describing the equilibrium plasma.
If one inserts a complete set of multi-particle states $|f\rangle$ between
the currents, and works in a basis $|i\rangle$ where ${\cal P}$ is diagonal,
then the correlator $W_{\mu\nu}$ can also be written as
\begin {equation}
    W_{\mu\nu}(K) = \int d^4x \> e^{-i K x} \sum_i {\cal P}_i \,
    \langle i | j_\mu(0) | f \rangle \,
    \langle f | j_\nu(x) | i \rangle .
\label{eq:Wmunu2}
\end {equation}
It will be useful to remember that this corresponds to scattering
from an initial state $i$ to a final state $f$ plus the emitted photon.
Moreover, the integrand in (\ref{eq:Wmunu2}) can be interpreted as
representing interference
between photon emission at the space-time points $0$ and $x$.

\begin{figure}[t]
\includegraphics[scale=0.45]{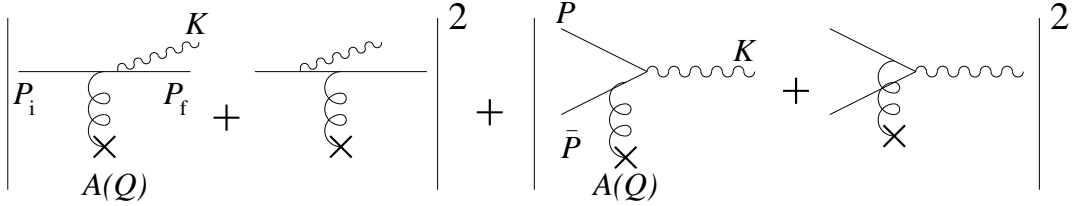}
\medskip
\caption{\label{fig:intro_classical}
  The perturbative \brem\ and annihilation processes of
  Fig.\ \ref{fig:inelastic}, with the soft gluon fields now interpreted as
  classical background fields.
}
\end{figure}

\begin{figure}[t]
\includegraphics[scale=0.6]{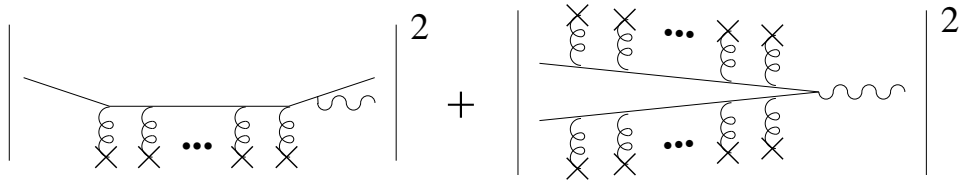}
\smallskip
\caption{\label{fig:multiple}
  Bremsstrahlung and annihilation processes including 
  the multiple interactions with a background field,
  which lead to the LPM effect.
}
\end{figure}

Our first approximation will be to treat the soft $(q \sim gT)$ gluons
of the problem as a random, {\it classical}\/ non-Abelian background field
$A_\mu(x)$, through
which the particles that \brem\ or annihilate propagate.
The lowest-order processes of Fig.\ \ref{fig:inelastic} are then replaced
by Fig.\ \ref{fig:intro_classical}, and our task will be to sum up
multiple gluon interactions such as depicted in Fig.\ \ref{fig:multiple}.
After computing rates, we will appropriately average over this
random classical gauge field.
Such averaging will be denoted by $\dlangle \cdots \drangle$.
By translation invariance, the variance of the background field
must have the form%
\footnote
    {%
    A translationally invariant choice of gauge fixing,
    such as Lorentz or Coulomb gauge, is tacitly assumed.
    }
\begin {equation}
   \dlangle A^\mu(Q)^*\,A^\nu(Q') \drangle
   = \rho^{\mu\nu}_A(Q) \; (2\pi)^4 \delta^{(4)}(Q{-}Q') ,
\label {eq:AA}
\label {eq:AAcorrelation}
\end {equation}
where $\rho^{\mu\nu}_A(Q)$ is the spectral density of thermal gauge field
fluctuations.
We may treat the statistical distribution of the background gauge field 
as Gaussian, dropping higher-order connected correlations.
This is equivalent to neglecting the nonlinear self-interaction of the
background and neglecting the interactions between different scatterers.
This approximation is valid, at leading order,
everywhere except in the deep infrared ($Q \sim g^2 T$).
Neglecting non-Gaussian correlations in the background gauge field would not
be allowable if the processes under consideration were sensitive
(at leading order) to ultra-soft $g^2 T$ fluctuations.
Fortunately, this is not the case \cite{paper1}.

Physically, these soft gauge fields are created by other charge carriers
[with typical $O(T)$ momenta]
passing randomly through the plasma.
The statistics of the soft gauge fields generated by these particles
can be described using
(i) the fluctuation-dissipation theorem, and
(ii) the standard hard thermal loop (HTL) approximation for
the retarded self-energy $\Pi(Q)$ of soft ($Q \ll T$) gauge fields.
Namely, the spectral density $\rho_A(Q) \equiv ||\rho_A^{\mu\nu}(Q)||$
may be taken to be
\begin {eqnarray}
   \rho_A(Q)
   & \simeq&
   2 [n_{\rm b}(q^0)+1] \, \Im\, G^{\rm HTL}_\ret(Q)
\nonumber\\
   & =&
   2 [n_{\rm b}(q^0)+1] \;
   \Im \left[
        \frac {1}{G^{(0)}_\ret(Q)^{-1} + \Pi^{\rm HTL}_\ret(Q)}
      \right] ,
\label {eq:rhoA}
\end {eqnarray}
where $n_{\rm b}(\omega) = 1/(e^{\beta\omega}{-}1)$ is the Bose distribution
function,
$G^{(0)}_\ret(Q)$ is the free retarded gauge field propagator,
and $G^{\rm HTL}_\ret(Q)$ is the retarded propagator with hard thermal loops
resummed, which account for Debye screening and Landau damping of
the color fields.
We will review the specific formulas later, in section \ref{sec:final}.
Because the background fields of interest are soft, one may
replace the $[\nb(q^0){+}1]$ statistical factor by
its low frequency limit, namely $T/q^0$.

The presence of correlations in the background field with non-trivial
frequency (and wave-number) dependence is the
most significant difference between our problem of hard photon emission
in a relativistic plasma, and the original setting for the LPM effect
of hard photon emission in a static medium of random Coulomb scatterers
(nuclei).  In order to be able to make contact later between our
results and those of Migdal, we will treat the spectral density
$\rho_A(Q)$ as arbitrary, and not use the specific form (\ref {eq:rhoA})
until the very end.

% ---------------------------------------------------------------------------

\subsection {Particle propagation in a random field}

Our next task is to rewrite the current correlator (\ref{eq:Wmunu})
in terms of propagation amplitudes of single particle states in the random
background.
We will focus on the \brem\ process and come back to pair
annihilation in section \ref{sec:pair}.
Bremsstrahlung corresponds
to a contribution to the current correlator (\ref{eq:Wmunu}) of the
form%
\footnote
    {%
    For any three-momentum $\p$,
    $\int_\p$ means $\int d^3\p / (2\pi)^3$,
    while for a four-vector $Q$,
    $\int_Q$ will denote $\int d^4Q / (2\pi)^4$.
    }
\begin {equation}
     W_{\mu\nu}^{\rm brem}(K) 
     =
	 \int d^4x \> e^{-iKx} \int_{\p_i\p_f}
         n(p_i) [1\pm n(p_f)] \>
   \Biggdlangle
	 \langle \p_i | j_\mu(0) | \p_f \rangle
              \> \langle \p_f | j_\nu (x) | \p_i \rangle
   \Biggdrangle \,,
\label {eq:Wbit}
\end {equation}
where $|\p_i\rangle$ and $|\p_f\rangle$ represent one-particle
states of the
emitting charge carrier, and
$n$ is the corresponding equilibrium distribution function
[either $n_{\rm b}(p)$ or $n_{\rm f}(p)$].
The contributions (\ref{eq:Wbit}) to the current correlator have
to be summed over all types (including anti-particles) and spins of
hard charged quasi-particles.
The momenta $\p_i$ and $\p_f$ can be regarded as the initial
and final momenta of the charge carrier,
and $n(p_i)$ and $1\pm n(p_f)$ are the initial and final state
statistical factors.
There is an approximation hiding here because momentum eigenstates
$|\p\rangle$ are not exact energy eigenstates in the presence of
the soft background field.
The density matrix is not therefore precisely $n(p)$.  Over the
$O(1/g^2 T)$ formation time relevant to our problem,
soft momenta transfers change
$\p$ by order $gT \ll p$.  But the resulting error in the use of
$n(p)$ is then sub-leading in $g$ and so can be ignored in a
leading-order analysis.

Making the time evolution explicit, and inserting complete sets of
intermediate states,
the statistically averaged matrix elements appearing in
(\ref{eq:Wbit}) can be rewritten as
\begin {eqnarray}
   && {} {\phantom{=}} \int_{\p_i'\p_f'}
    \Biggdlangle
      \, \langle \p_i | j_\mu(0) | \p_f \rangle
      \, \langle \p_f | U(0,x^0) | \p_f' \rangle
      \, \langle \p_f' |j_\nu(\x) | \p_i' \rangle
         \langle \p_i' | U(x^0,0) | \p_i \rangle
   \Biggdrangle
\nonumber\\ && {}
   = \int_{\p_i'\p_f'}
                 \langle \p_f | j_\mu(0) | \p_i \rangle^*
              \, \langle \p_f' |j_\nu(\x) | \p_i' \rangle
              \, \biggdlangle
                   \langle \p_f' | U(x^0,0) | \p_f \rangle^* \,
                   \langle \p_i' | U(x^0,0) | \p_i \rangle
              \biggdrangle \,,
\label {eq:Wbrem1}
\end {eqnarray}
where $U(t',t)$ is the time evolution operator in the background field.
As usual, the spatial Fourier transform in (\ref {eq:Wbit}) combines
with translation invariance (of the statistically averaged matrix elements)
to enforce total momentum conservation:
\begin {equation}
   \k = \p_i - \p_f = \p_i'-\p_f' \,.
\end {equation}

\begin{figure}[t]
\includegraphics[scale=0.70]{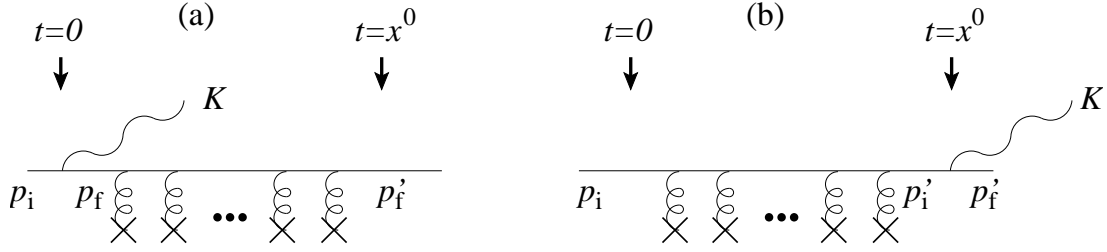}
\medskip
\caption{\label{fig:bremx}
   Two diagrams, time ordered from left to right, whose interference
   contributes to the rate of \brem.  The first diagram represents
   photon emission at time zero, and the second at time $x^0$, and
   in both cases the diagrams show the evolution between these two times.
}
\end{figure}

\begin{figure}[t]
\includegraphics[scale=0.70]{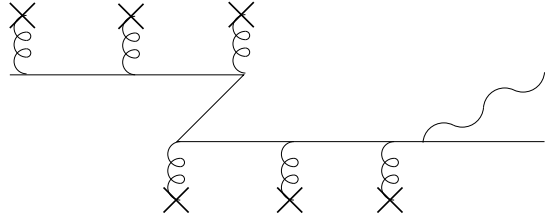}
\caption{\label{fig:Z}
   A time-ordered Z diagram.
}
\end{figure}

\begin{figure}[t]
\bigskip
\includegraphics[scale=0.70]{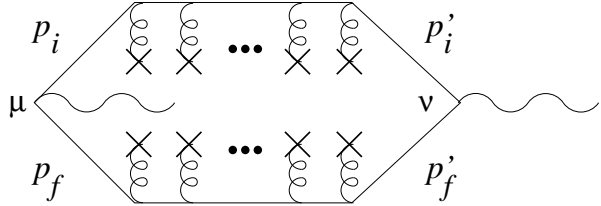}
\bigskip
\caption{\label{fig:jj}
   A single diagram depicting the interference of the two diagrams
   of Fig.\ \ref{fig:bremx}.  The interactions along top and bottom lines
   are independently time ordered from left to right.
}
\end{figure}

Diagrammatically, this contribution represents the interference in
the evolution of the particle from time 0 to $x^0$ depending on whether
the photon is emitted at time 0 or at time $x^0$, as depicted in
Figs.\ \ref{fig:bremx}a and b.  These diagrams can be considered
as time ordered (with time running from left to right)
because each individual $Q \sim gT$
momentum transfer is not enough to create or destroy a
(nearly on-shell) particle/anti-particle pair.
That is, in a
time-ordered Z contribution like Fig.\ \ref{fig:Z}, the three-particle
intermediate state would have to be so far off shell that its contribution is
suppressed compared to the time ordering of Fig.\ \ref{fig:bremx}a.
It is convenient to put the interference of the evolutions of
Figs.\ \ref{fig:bremx} together into the single diagram depicted
by Fig.\ \ref{fig:jj}.  The top line (or ``rail'') represents
$\langle \p_i' | U(x^0,0) | \p_i \rangle$ and the bottom line (rail) the
complex conjugate of $\langle \p_f' | U(x^0,0) | \p_f \rangle$.
This looks just like a Feynman diagram for the current correlator,
except with the added interpretation that each rail of the
diagram can be considered as time ordered from left to right.

% ---------------------------------------------------------------------------

\subsection {Ordering of interactions}

The next step is to understand the effect of averaging over the background
gauge field.  A quick way to visualize the dominant correlations is
to think of the photon and emitter qualitatively as approximately classical
particles,
having both approximately well-defined position and momentum, obeying
roughly
\begin {equation}
   \x(t) \simeq \x(0) + \v_\p \, t \, .
\end {equation}
Indeed, the original, qualitative derivation of the LPM effect was
purely classical \cite{LP}.%
\footnote{
   Classical results become precise quantitatively in the limit that
   the photon energy is small compared to the emitter energy.
}
(See section 1 of Ref.\ \cite{gyulassy&wang} for a concise review.)
More formally, instead of using a basis of momentum or position states,
one may use an (over-complete) basis of Gaussian wave packets
with width
$\Delta x \sim (\Delta p)^{-1} \sim (gT)^{-1}$.  For momenta of order $T$,
the widths will remain of this order over the time scale $1/g^2 T$
relevant to our problem.%
\footnote{
   In more detail, the transverse momenta in the wave packet will be order
   $\Delta p_\perp \sim \Delta p \sim gT$,
   both from the initial width as well as momentum transfer from scattering.
   The transverse spatial spread of the wave packet
   over time $\tau \sim 1/g^2 T$ will then be
   $\tau \Delta p_\perp/p \sim 1/gT$, since
   $p \sim T$.  There is also a longitudinal spread in the wave packet
   due to the effective thermal mass $m_\infty \sim gT$ of hard quarks,
   which gives a dispersion in velocities of order
   $\Delta v \sim \Delta (p/\sqrt{p^2+m_\infty^2})
   \sim m_\infty^2 \Delta p / p^3 \sim g^3$.  The longitudinal spread of
   the wave packet due to this dispersion is then of order
   $\tau \, \Delta v \sim g/T$.
}
The spatial uncertainty $\Delta x$ in the position
of the particles is then always small
compared to $1/g^2 T$, which is the mean free path for the soft gluon
interactions with $Q \sim gT$ which dominate our problem.
This implies that interactions with scatterers must
occur in a definite order, regardless of when the \brem\ photon is
emitted.

\begin{figure}[t]
\includegraphics[scale=0.70]{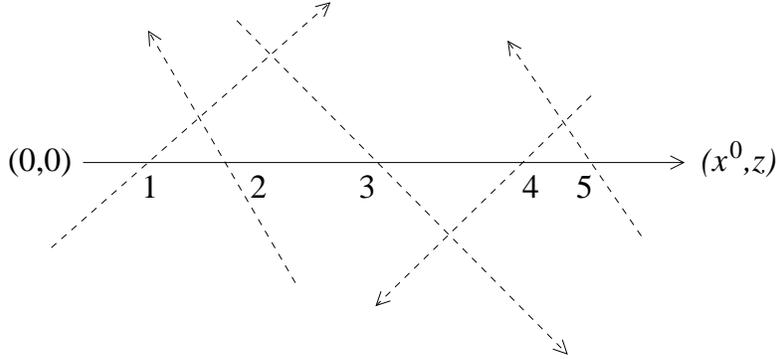}
\caption{\label{fig:fuzzy}
   The emitter's trajectory, thought of as approximately classical,
   and the ordering in time and space this gives to its local interactions
   with scatterers (dashed lines).
}
\end{figure}

\begin{figure}[t]
\includegraphics[scale=0.60]{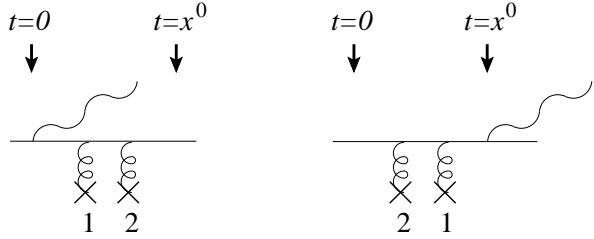}
\medskip
\caption{\label{fig:disorder}
   Two amplitudes which do not interfere (at leading order) because of
   mismatched order of encountering scatterers.
}
\end{figure}

\begin{figure}[t]
\includegraphics[scale=0.60]{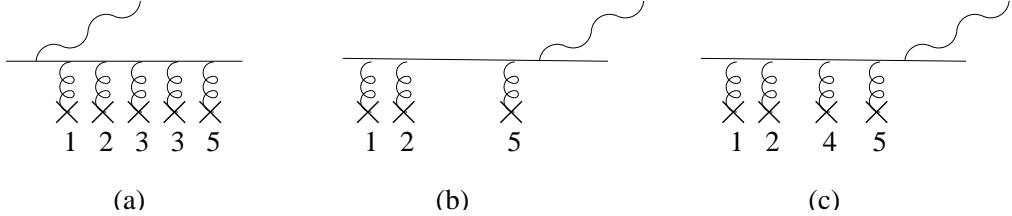}
\caption{\label{fig:order}
   Amplitudes for which the time and space
   ordering of interactions is consistent.
}
\end{figure}

\begin{figure}[t]
\includegraphics[scale=0.60]{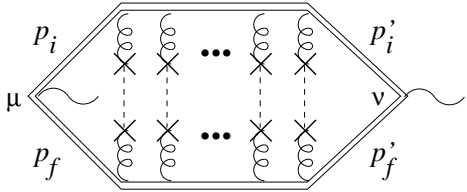}
\medskip
\caption{\label{fig:jj2}
   The product of amplitudes, as in Fig.\ \ref{fig:jj},
   now averaged over random background fields.
   The dashed lines denote the independent correlations
   $\dlangle A A \drangle$ of the background gauge fields.
   The double lines represent the resummed propagators
   of Fig.\ \ref{fig:resum}.
}
\end{figure}

\begin{figure}[t]
\medskip
\includegraphics[scale=0.60]{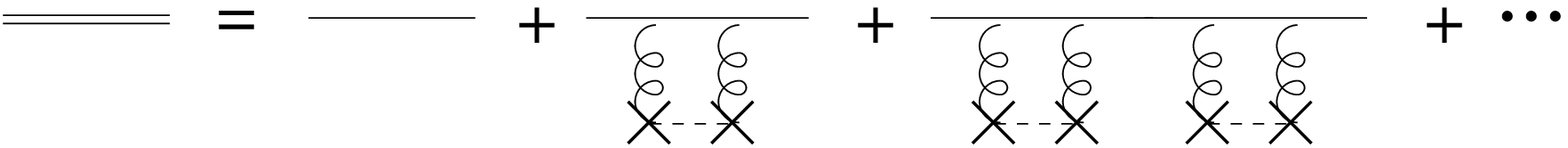}
\caption{\label{fig:resum}
   The resummation of self-energy insertions into the propagator.
}
\end{figure}

Fig.\ \ref{fig:fuzzy} illustrates the basic point.
It shows the emitter propagating along the $+z$ axis as a (nearly)
localized classical particle.
The dashed lines depict the small subset of other hard particles in
the plasma which happen to pass close to the emitter during its flight
and happen to interact with it when they do.
Because of the localization of the emitter in space and time,
and because of the local nature of the soft interactions
(set by the Debye screening length $1/gT$),
the interactions (for the particular background of
hard particles shown) must happen in the order numbered in the diagram,
regardless of when the collinear photon is emitted.
That is, there can be no interference
(at leading order in $g$) between the diagrams of
Fig.\ \ref{fig:disorder} if the numbers 1 and 2
denote the fields generated by two different scatterers.
However, the diagrams of Fig.\ \ref{fig:order}a and b could interfere.
So could the diagrams of Fig.\ \ref{fig:order}a and c, except that
this particular interference vanishes when averaged over the randomness of
the scatterers: the unmatched scatterer 4 could be
a particle or anti-particle (for example),
and a single insertion of its field in
the product of amplitudes averages to zero by charge conjugation invariance.
So, after averaging, non-zero contributions only arise from the cases where
interactions with a given scatterer happen to appear twice in
Fig.\ \ref{fig:jj}, with the interactions of the top and bottom rails
consistently ordered in time and space.
We can rewrite the random average of these
figures in the form of Figs.\ \ref{fig:jj2} and \ref{fig:resum},
where the dashed lines
indicate that the corresponding pair of $A$'s have been replaced
by the correlation $\dlangle A_\mu A_\nu \drangle$.
In our problem, this correlator is given by (\ref{eq:AAcorrelation}),
but the argument we have used
to arrive at Fig.\ \ref{fig:jj2} would also apply to {\it static}\/ random
scatterers, such as considered by Migdal, whenever one has a similar hierarchy
of scales.

We should mention in passing
that the condition that the entire wave packet remain well-localized,
in all three dimensions, on a scale parametrically small compared to
the mean separation between scattering events,
is much stronger than necessary to argue that the interaction
must be ordered in time (at leading order).
Interactions are effectively ordered in time, and ladder diagrams dominate,
in numerous other applications including the classic case of scattering
from point-like static random impurities.%
\footnote
    {%
    In this case,
    a localized wave packet impinging on a single scattering center
    will produce a scattered wave which resembles an outgoing spherical shell,
    with no localization in direction whatsoever.
    But as long as the temporal duration of the scattered wave
    (as it moves past a fixed location) is small compared to
    the mean time between collisions, multiply-scattered waves
    produced by scattering, in different orders, off a given set of
    scattering centers, will have negligible overlap and hence not interfere.
    In particle-oriented language, all we are saying is that if a
    particle travels from some localized source
    to a localized detector by scattering twice,
    first off impurity $A$ and then off $B$,
    its path length will be different than if it scatters first
    off $B$ and then off $A$.
    As long as this difference in path length is large compared
    to the spatial extent of the wave packet, no significant
    interference can occur.
    The basic criterion which is relevant here, and in all applications
    of kinetic theory, is that the mean time between collisions must be
    large compared to both the temporal duration of the initial wave packet
    (or $\hbar$ over its energy), and to the time duration of a single
    scattering event (as determined by the energy derivative of the scattering
    amplitude).
    }

The self-energy resummation on the top and bottom rails,
involving insertions of the background field correlator
as shown in Fig.~\ref {fig:resum},
represents the inclusion of the thermal width $\Gamma$
in the propagator of the emitting particle.
In hot gauge theories, the thermal width of quasiparticles is dominated
by soft scattering and is actually an
infrared divergent quantity in perturbation theory.
However, these divergences (which represent sensitivity to ultra-soft
fluctuations) will turn out to cancel in our final result
when combined with the correlations that connect the top and bottom rails.
Our final formulas will be well behaved in the infrared.

Before moving on to the evaluation of the ladder diagrams of
Fig.\ \ref{fig:jj2}, we may use the fuzzy classical particle picture
to understand why the LPM effect depends crucially on the fact that the
photon and the emitting particle move at or very close to the speed of
light.
If a particle of momentum $p$ moving in the $+z$ direction
emits a collinear photon of momentum $k$ at time and position zero,
then the subsequent positions of the particle
and the photon at time $t'$ will be
\begin {eqnarray}
   z_{\rm photon} &\simeq& v_\gamma \, t' \,,
\qquad
   z_{\rm particle} \simeq v_{p-k} \, t' \,,
\end {eqnarray}
where $v_\gamma$ is the velocity of the photon
and $v_p$ the velocity of the (quasi) particle with momentum $p$.
If, on the other hand,
the same particle doesn't emit the photon until time $t$,
the subsequent positions will be
\begin {eqnarray}
   z_{\rm photon} \simeq v_p \, t + v_\gamma \, (t'-t) \,,
\qquad
   z_{\rm particle} \simeq v_p \, t + v_{p-k} \, (t'-t) \,.
\end {eqnarray}
These two processes can interfere significantly only if the photon and
particle trajectories are the same, up to the fuzziness $\Delta x$.
That requires
\begin {equation}
   v_\gamma t \simeq v_p t \simeq v_{p-k} t ,
\end {equation}
which is natural (that is, not suppressed by additional phase space
restrictions) only if the particles are all ultrarelativistic.

% ===========================================================================

\section {Ladder diagrams as integral equations}
\label {sec:integral}

\subsection {Relativistic Schr\"odinger equation}

A concise and convenient method for representing the propagation of
nearly on-shell particles in a soft gauge background is with the
relativistic one-particle Schr\"odinger equation.  The free equation is
\begin {equation}
   i \partial_t \psi = E_p \psi ,
\label {eq:QM}
\end {equation}
where $E_p = \sqrt{p^2+m^2}$ and
$m$ is the appropriate mass (in our case $m_\infty$, which
characterizes the in-medium dispersion correction of a fermion of
negligible explicit mass).
The corresponding retarded free propagator is just
\begin {equation}
   G_0(P) = \frac{i}{p^0 - E_p + i \epsilon} \,,
\label {eq:GQM}
\end {equation}
and satisfies
\begin {equation}
    \theta(t{-}t') \>
    \langle \p | U_0(t,t') | \p' \rangle
    =
    \int {dp^0 \over 2\pi} \> e^{-i p^0 (t-t')} \>
    G_0(P) \; \langle \p | \p' \rangle \,.
\end {equation}
We will use the traditional
normalization of one-particle quantum mechanics,
\begin {equation}
   \langle \p | \p' \rangle = (2\pi)^3 \, \delta^{(3)}(\p{-}\p') \,,
\label {eq:QMnorm}
\end {equation}
as opposed to the usual convention in relativistic field theory, which
has an extra factor of $2 E_p$ on the right hand side.
Except for the change in normalization, $G_0(P)$ is just the positive-energy
pole piece of the usual propagators of relativistic field theory
(or, equivalently, the energy denominator of time-ordered perturbation theory
when only single particle intermediate states contribute).
The advantage of the description used here is that it does not depend on
the spin or type of particle; it can be used to economically describe the
propagation of hard scalars, fermions, or gauge bosons in soft background
fields.

In a soft background gauge field
(containing fluctuations on a scale $Q \sim gT \ll T \sim P$),
the single-gluon vertex is just
\begin {equation}
   i g \, v \cdot A \,,
\label {eq:QMvertex}
\end {equation}
up to corrections suppressed by $Q/P \sim g$,
where $v$ is defined by
\begin{equation}
   v \equiv \frac{P}{p^0} = (1,\v) \simeq (1,\hat\p) \,.
\end {equation}
A simple, quick way to see this is to replace
$P$ by $P-gA$ in the Schr\"odinger equation (\ref{eq:QM}) and
propagator (\ref{eq:GQM}):
\begin {equation}
   i \partial_t \, \psi - g A_0 = E_{\p-g{\bf A}} \, \psi .
\end {equation}
Because the background field is soft,
one may treat $A$ and $P$ as if they commute, since commutators will replace
the operator $P$, whose relevant matrix elements are $O(T)$,
by $Q\sim gT$, and hence only generate a subleading contribution.
Expanding the equation
to first order in $A$, one then obtains
\begin {equation}
   i \partial_t \psi = (E_p - g v\cdot A) \, \psi + O(A^2) ,
\end {equation}
which gives the interaction (\ref{eq:QMvertex}).
Only the ``convective'' contribution of the charged particle
to the gauge current appears in the soft gluon vertex (\ref {eq:QMvertex});
the spin dependent contributions which would normally distinguish
scalars, fermions, or gauge bosons are absent.
Spin dependent contributions to the current are suppressed
by an additional power of $Q/P \sim g$, and hence only generate
sub-leading corrections to the propagation of a hard excitation
through the soft background gauge field.

{}From the point of view of one-particle quantum mechanics,
anti-particles are just another type of particle.
The propagation of nearly on-shell anti-particles in a soft
background field is also described by the propagator (\ref{eq:GQM})
and vertex (\ref{eq:QMvertex}) except that the gauge field $A$ should
be in the conjugate representation.
Readers who prefer a discussion in terms of standard
relativistic Feynman rules may refer to Ref.\ \cite{paper1}.

% ---------------------------------------------------------------------------

\subsection {The integral equation}

The contribution to the leading-order photon emission rate from
the sum of the ladder diagrams depicted in Fig.~\ref{fig:jj2}
may be expressed in terms of the solution to a suitable integral equation.
Analogous summations of ladder diagrams are, of course,
well known in many other contexts
(such as the Bethe-Salpeter equation for bound states),
although the form of the resulting integral equation
intimately depends on the precise form of the propagators making up the
rungs and side rails of the ladder diagrams, as well as the specific
vertex factors at the ends of the diagrams.

The group structure of the diagrams of Fig.\ \ref{fig:jj2} is trivial:
for every explicit power of $g^2$, there is also one factor of the quadratic
Casimir $C_R$ for the color representation $R$ of the emitter.
This Casimir is defined by $T_R^a T_R^a = C_R$ where
$T_R^a$ are the generators of the representation,
normalized so that ${\rm tr} \, T_R^a T_R^b = \half \, \delta^{ab}$.
For conventional quarks or anti-quarks in the fundamental representation
of color SU(3), this Casimir is
\begin {equation}
   C_R = \tfrac{4}{3} \quad \mbox{ [fundamental SU(3)]} \,.
\end {equation}

Our first step will be to resum the self-energy insertions of
Fig.\ \ref{fig:resum} by switching from the propagator (\ref{eq:GQM})
to the resummed propagator
\begin {equation}
   \frac{i}{p^0 - E_\p \pm i\Gamma_\p} ,
\end {equation}
where the $+$ sign gives the retarded propagator
and the $-$ sign the advanced one.
The width $\Gamma_\p$ is proportional to the imaginary part of the
self-energy of the hard particle.
The explicit form for this width is discussed below.
The corresponding real part of the self energy is absorbed into $E_\p$.
In fact the real part, unlike the imaginary part,
is dominated by hard rather than soft collisions and cannot
be directly computed in the soft background formalism we are using.
Instead, it can be taken directly from the well-known hard thermal loop result,
and just gives the usual effective thermal mass.
For hard quarks, the effective energy is then
\begin {equation}
   E_\p = \sqrt{p^2+m_\infty^2} ,
\end {equation}
up to yet higher-order corrections,
where \cite{Klimov,Weldon}
\begin {equation}
   m_\infty^2 = {\textstyle {1\over4}} \, C_R \, g^2 T^2 \,.
\end {equation}

\begin{figure}[t]
\includegraphics[scale=0.50]{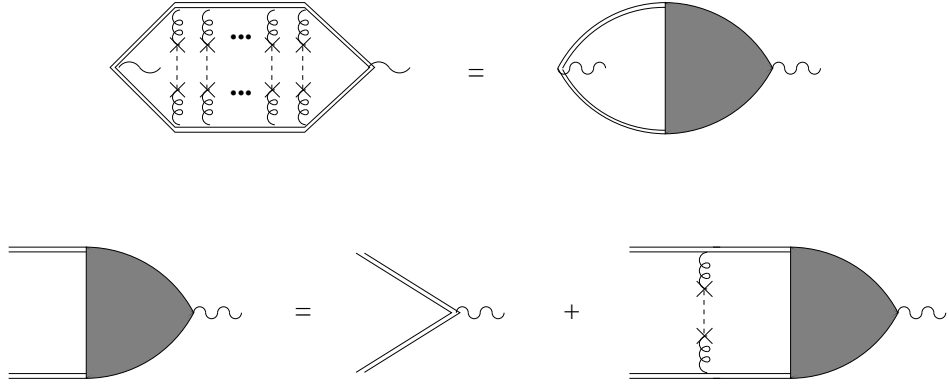}
\medskip
\caption{\label{fig:sd}
   The leading-order contributions to the current-current correlator
   $W_{\mu\nu}$ expressed in terms of the solution to
   a linear integral equation.
}
\end{figure}

The sum of interferences represented by the diagrams of
Fig.\ \ref{fig:jj2} may now be rewritten in terms of a linear integral
equation as depicted in Fig.\ \ref{fig:sd}.
To write this explicitly,
it is convenient to work in frequency and momentum space.
Consider the contribution to the 
correlator $W_{\mu\nu}$ from the $x^0 < 0$
portion of the $x^0$ integration in Eq.~(\ref{eq:Wbit}).%
\footnote
    {%
    We could equally well focus on the $x^0 > 0$ portion.
    As noted below, the two contributions are just complex conjugates
    of each other.
    It happens to be the $x^0 < 0$ portion which leads to the
    specific form of the integral equation derived in our earlier work
    \cite {paper1}.
    }
If we rewrite the time evolution matrix elements in Eq.~(\ref {eq:Wbrem1})
as
$
  \dlangle
       \langle \p_f | U(0,x^0) | \p_f' \rangle \,
       \langle \p_i | U(0,x^0) | \p_i' \rangle^*
  \drangle
$
then we may use retarded propagators in evaluating these matrix elements.
The resulting contribution to the current correlator,
illustrated in Fig.\ \ref{fig:sd}, is
\begin {equation}
  \int_P
        n(\p{+}\k) \, [1\pm n(\p)] \,
        \langle \p | j_\mu(\k) | \p{+}\k \rangle^* \,
        \fff_\nu(P;K) ,
\end {equation}    
where $\fff_\nu(P;K)$, which represents the shaded vertex
in Fig.\ \ref{fig:sd}, satisfies the integral equation%
\footnote{
   $\fff_\nu$ here is analogous to the combination ${\cal F}{\cal D}_\nu$ of
   Ref.\ \cite{paper1}.  We have chosen our conventions for
   defining $\fff_\nu$ so that later equations will closely resemble
   those of Refs.\ \cite{paper1,paper2}.
}
\begin {eqnarray}
    \fff_\nu(P;K) &=&
    \left( \frac{i}{(p^0+k^0) - E_{\p+\k} + i \Gamma_{\p+\k}} \right)^*
    \frac{i}{p^0 - E_\p + i \Gamma_\p}
\label {eq:SD1}
\\ && {} \times
    \left[
	\vphantom{{1\over E_\p}}
        \langle \p | j_\nu(\k) | \p{+}\k \rangle
        + g^2 C_R \int_Q \biggdlangle
                     [v_{\p{+}\k}\cdot A(Q)] \, [v_\p\cdot A(Q)]^* 
                 \biggdrangle \,
                 \fff_\nu(P{-}Q;K)
    \right] .
\nonumber
\end {eqnarray}
The color matrices and the momentum-conserving delta function
$(2\pi)^4\delta^{(4)}(\cdots)$
are to be understood as already factored out of the
$\dlangle A A \drangle$ correlation in this (and subsequent) formulae.
The $x^0 > 0$ region of the $x^0$ integration in (\ref{eq:Wbit}) simply
gives the complex conjugate (the replacement of retarded by advanced
propagators), so that the total \brem\ contribution, from
a single carrier type and spin state, is
\begin {equation}
  W_{\mu\nu}^{\rm brem}(K) = 2 \, \Re \left[\int_P
        n(\p{+}\k) \, [1\pm n(\p)] \,
        \langle \p | j_\mu(\k) | \p{+}\k \rangle^* \,
        \fff_\nu(P;K) \right]\,.
\label {eq:SD1w}
\end {equation}

Note that the photon four-momentum $K$ is fixed,
but the integral equation relates $\fff_\nu(P;K)$ at
different values of the four-vector $P$.
In fact, at the order of interest, the dependence on $P$ may be simplified.
Let $\ppar$ represent the component of $\p$ in the direction $\k$
of the photon,
and let $\p_\perp$ be the part of $\p$ perpendicular to $\k$.
In our process, the momentum transfer is
$O(gT)$, and so the relevant values of $\p_\perp$ are $O(gT)$.
Expanding in both $\p_\perp$ and the thermal mass $m_\infty$,
one has
\begin {eqnarray}
   p^0 - E_\p + i \Gamma_\p
   &\simeq& (p^0-\ppar) - \frac{p_\perp^2+m_\infty^2}{2 \ppar}
          + i \Gamma_\p \,,
\\
   (p^0+k^0) - E_{\p+\k} + i \Gamma_{\p+\k}
   &\simeq& (p^0-\ppar) - \frac{p_\perp^2+m_\infty^2}{2 (\ppar+k)}
          + i \Gamma_{\p+\k} \,,
\end {eqnarray}
where we have assumed that the photon is on-shell, and that
$\ppar > 0$ so that the photon is traveling in the same
direction as the emitter.
The last two terms in both equations are $O(g^2 T)$, and so each of the
propagators will be suppressed unless $p^0{-}\ppar$ is
also $O(g^2 T)$.%
\footnote
    {
    If $\ppar < 0$, then the two propagators are peaked near
    $p^0 = -\ppar$ and $p^0 + k = |\ppar + k|$,
    and it is impossible for both propagators
    to simultaneously be (nearly) on-shell.
    Hence, only the region $\ppar > 0$ contributes
    to the leading-order \brem\ rate.
    }
If ${\cal S}_\nu(P;K)$ is only ever going to be integrated against functions
that are smooth on the scale of $O(g^2 T)$ variations in $p^0{-}\ppar$,
then one may approximate
\begin {eqnarray}
    \left( \frac{i}{(p^0+k^0) - E_{\p+\k} + i \Gamma_{\p+\k}} \right)^*
    \frac{i}{p^0 - E_\p + i \Gamma_\p}
    &\simeq&
    \frac{2\pi i\>\delta(p^0{-}\ppar) \, \theta(\ppar)}{
        (E_{\p+\k}+i\Gamma_{\p+\k})-(k^0+E_\p-i\Gamma_\p)}_{\strut}
\nonumber\\
    &\simeq&
    \frac{2\pi \>\delta(p^0{-}\ppar) \, \theta(\ppar)}{
        i\,\delta E + (\Gamma_{\p+\k}+\Gamma_\p)} \,,
\end {eqnarray}
where
\begin {equation}
   \delta E \equiv \frac{ k \, (p_\perp^2+m_\infty^2) }{ 2\ppar \, (k+\ppar) }
   \simeq
   k^0+E_\p-E_{\p+\k} \,.
\label {eq:dEdef}
\end {equation}
This is commonly known as the pinching pole approximation, and in
our analysis is justified for computing the rate at leading order in $g$.
Making this substitution into the integral equation (\ref{eq:SD1}),
one sees that the resulting solution will have the form
\begin {equation}
   \fff_\nu(P;K) = \ff_\nu(\p;K) \> 2\pi\,\delta(p^0{-}\ppar) \, \theta(\ppar)
   \,.
\end {equation}
Inserting this form, and rewriting
\begin {equation}
   \delta(p^0{-}\ppar) \, \delta\blparen(p^0{-}q^0)-(\ppar{-}\qpar)\brparen
   = \delta(p^0{-}\ppar) \, \delta(q^0{-}\qpar) \,,
\end {equation}
one finds
\begin {equation}
  W_{\mu\nu}^{\rm brem}(K) = 2 \, \Re \int_\p
        n(\p{+}\k) \, [1\pm n(\p)] \,
        \langle \p | j_\mu(\k) | \p{+}\k \rangle^* \>
        \ff_\nu(\p;\k) \> \theta(\ppar) ,
\label {eq:SD2w}
\end {equation}    
where
\begin {eqnarray}
    \ff_\nu(\p;\k) &=&
    \frac{1}{i\delta E + (\Gamma_{\p+\k}+\Gamma_\p)}
    \Biggl[
        \langle \p | j_\nu(\k) | \p{+}\k \rangle
\nonumber
\\ && {}
        + g^2 C_R \int_Q 2\pi\,\delta(q^0{-}\qpar) \, \biggdlangle
                     [v_{\p+\k}\cdot A(Q)] \, [v_\p\cdot A(Q)]^* 
                 \biggdrangle \,
                 \ff_\nu(\p{-}\q;\k)
    \Biggr] .
\label {eq:SD2}
\end {eqnarray}
In terms of the current matrix elements appearing in
the original definition (\ref {eq:Wbit}),
the function $\ff_\nu(\p;\k)$ is given by that
portion of the time-evolved off-diagonal matrix element of the current,
evaluated in the fluctuating background gauge field,
which is phase-coherent with the emitted photon.
That is, to leading order in $\gs$,
\begin {equation}
    \int d^4 x \> \theta(-x^0) \> e^{-i K x} \>
    \Bigdlangle \langle \p | j_\nu(x) | \p{+}\k \rangle \Bigdrangle
    =
    \ff_\nu(\p;\k) \, \theta(\ppar) \,.
\label {eq:jtoS}
\end {equation}
Time evolution is, of course, implicit in
$j_\nu(x) = j_\nu(x^0,\x) = U(0,x^0)\,j_\nu(\x)\,U(x^0,0)$.

In the interaction term of Eq.~(\ref{eq:SD2})
one may, at leading order, replace $v_\p \cdot A$ by
\begin {equation}
    A^+ \equiv A^0 - A_\parallel
\end {equation}
and rewrite the integral equation in the form
\begin {eqnarray}
    \langle \p | j_\nu(\k) | \p{+}\k \rangle
    &=&
    [i \, \delta E + (\Gamma_{\p+\k}+\Gamma_{\p})] \, \ff_\nu(\p;\k)
\nonumber
\\ && {}
        - g^2 C_R \int_Q 2\pi\,\delta(q^0{-}\qpar) \, \Bigdlangle
                     A^+(Q) \, [A^+(Q)]^* 
                 \Bigdrangle \,
                 \ff_\nu(\p{-}\q;\k)
    \,.
\label {eq:SD3}
\end {eqnarray}
As it stands, this integral equation mixes together different values of $\ppar$
(as well as different $\p_\perp$) 
in the argument of $S(\p;\k)$, due to the $O(gT)$ momentum
transfers in the interaction term.
The characteristic size of $\ppar$ for a hard quasiparticle is $O(T)$,
and no elements in the equation are sensitive (at leading order)
to $gT$ variations in $\ppar$.
Consequently,
we can treat $\ppar$ as fixed inside the interaction term,%
\footnote{
   This may be argued in more detail as follows.
   Given the linearity of the integral equation,
   one could multiply the source $\langle\p|j_\nu(\k)|\p{+}\k\rangle$ in
   (\ref{eq:SD3}) by $\delta(\ppar{-}\ppar')$, at the cost of introducing an
   integral over $\ppar'$ in the final expression (\ref{eq:SD2w}).
   The resulting solution to the integral equation would then have
   its dominant support on values of
   $\ppar$ which differ from $\ppar'$ 
   only by total momentum transfers of $O(gT)$.
   Because the
   $\langle \p | j_\nu(\k) | \p{+}\k \rangle^* \> \theta(\ppar)$
   factor in (\ref{eq:SD2w}) is smooth in $\ppar$
   (for $\ppar$ of order $T$),
   no error is made at leading order if
   $\ppar$ is replaced by $\ppar'$ in these terms.
   If the modified (\ref{eq:SD3})
   is then integrated over $\ppar$, and $\int d\ppar \> S(\p;\ppar',\k)$
   is renamed $S(\p_\perp;\ppar',\k)$, one arrives at
   Eqs.\ (\ref{eq:SD2w}) and (\ref{eq:SD4}).
}
replacing the equation by
\begin {eqnarray}
    \langle \p | j_\nu(\k) | \p{+}\k \rangle
    &=&
    [i \, \delta E + (\Gamma_{\p+\k}+\Gamma_\p)] \, \ff_\nu(\p_\perp;\ppar,\k)
\nonumber
\\ && {}
        - g^2 C_R \int_Q 2\pi\,\delta(q^0{-}\qpar) \, \Bigdlangle
                     A^+(Q) \, [A^+(Q)]^* 
                 \Bigdrangle \,
                 \ff_\nu(\p_\perp{-}\q_\perp;\ppar,\k) \, .
\label {eq:SD4}
\end {eqnarray}
This is now a linear integral equation in $\p_\perp$ space,
for fixed values of $\ppar$ and $\k$.

One remaining complication is that both the width $\Gamma_\p$ 
and the total scattering cross section have infrared divergences in our
approximations, proportional to the $Q \to 0$ divergence of
$
\int_Q 2\pi \delta (q^0{-}\qpar) \> \dlangle A^+(Q) [A^+(Q)]^* \drangle
$.
This divergence of $\Gamma_\p$ in HTL perturbation theory is well known
\cite{Gamma_divergence},
arises from collisions via the exchange of unscreened low-frequency
magnetic gluons, and is a symptom of sensitivity to ultra-soft non-perturbative
magnetic physics at the momentum scale $g^2 T$.
But this apparent infrared sensitivity is actually illusory in our problem,
as may be seen by rewriting the width in a form similar to the
interaction term in Eq.~(\ref{eq:SD4}).  The width comes from the imaginary
parts of the self-energy insertions in Fig.\ \ref{fig:resum}
for nearly on-shell particles and, to leading order, is given by
\begin {eqnarray}
   \Gamma_\p
   &=& \Im \left[
        i g^2 C_R \int_Q \frac{i}{(p^0 {-} q^0) - E_{\p-\q} + i \eps} \,
        \Bigdlangle A^+(Q) \, A^+(-Q) \Bigdrangle
     \right]_{p^0=E_\p}
\nonumber\\
   &\simeq&
        g^2 C_R \int_Q \pi \delta(q^0 {-} \qpar) \,
        \Bigdlangle A^+(Q) \, [A^+(Q)]^* \Bigdrangle .
\label {eq:Gam}
\end {eqnarray}
Substituting this into (\ref{eq:SD4}), one obtains the infrared-safe equation
\begin {eqnarray}
    \langle \p | j_\nu(\k) | \p{+}\k \rangle
    &=&
    i \, \delta E \, \ff_\nu(\p_\perp;\ppar,\k)
        + g^2 C_R \int_Q 2\pi\,\delta(q^0{-}\qpar) \, \Bigdlangle
                     A^+(Q) \, [A^+(Q)]^* 
                 \Bigdrangle \,
\nonumber
\\ && \kern 1.9in {} \times
                 [\ff_\nu(\p_\perp;\ppar,\k)
                       -\ff_\nu(\p_\perp{-}\q_\perp;\ppar,\k)] \, .
\label {eq:SDbrem}
\end {eqnarray}
We may also tidy up the expression for $W_{\mu\nu}$ by noting that
$n(p) \simeq n(\ppar)$ [since $\p_\perp$ is $O(gT)$], so that to leading order
\begin {equation}
  W_{\mu\nu}^{\rm brem}(K) = 2 \, \Re \int_\p
        n(\ppar{+}k) \, [1\pm n(\ppar)] \>
        \langle \p | j_\mu(\k) | \p{+}\k \rangle^* \>
        \ff_\nu(\p_\perp;\ppar,\k) \> \theta(\ppar) \, .
\label {eq:SDbremw}
\end {equation}

\subsection {Current matrix elements}

The current matrix element $\langle \p | j_\mu(\k) | \p{+}\k \rangle$
is the only ingredient of our equations that depends on the nature
of the emitter ({\it i.e.}, scalar or fermion).
Since $\p_\perp$ is parametrically small compared to $\ppar$,
it is sufficient to evaluate these matrix elements in the limit
of small transverse momentum.
We only need the transverse components of $j_\mu$
because the
currents are dotted into photon polarization vectors in Eq.~(\ref{eq:Wmunu}).
By rotational invariance about the $\k$ axis, these must be proportional
to $\p_\perp$.
Let $q e$ denote the electric charge of the emitting particle
(so that up quarks, for example, have $q = 2/3$).
Then in the small $\p_\perp$ limit,
the transverse current matrix element takes the form
\begin {equation}
    \langle \p | \j_\perp(\k) | \p{+}\k \rangle
    \simeq 2 q e \, \p_\perp \, {\cal J}_{\ppar \leftarrow\ppar{+}k} ,
\label {eq:Jlimit}
\end {equation}
where we have chosen to factor out the charge of the emitter.
The explicit form of the ``splitting function''
${\cal J}_{\ppar \leftarrow\ppar{+}k}$,
which depends on $\ppar$ and $k$ as well as the spin of the emitter,
will be discussed momentarily.
Given the linearity of the integral equation (\ref {eq:SDbrem})
(and the fact that it only couples different values of $\p_\perp$, not $\ppar$),
we can factor out all dependence on the emitter type
by redefining the transverse part of $\ff_\nu(\p_\perp;\ppar,k)$ as
\begin {equation}
    {\bf S}_\perp(\p_\perp;\ppar,\k)
    = q e \, {\cal J}_{\ppar\leftarrow\ppar{+}k} \> {\bf f}(\p_\perp;\ppar,\k)
\end {equation}
to obtain
\begin {equation}
  \sum_a {\eps^{\mu}_{a}}^* \eps^\nu_{a} W_{\mu\nu}^{\rm brem}(K) =
        2 (qe)^2 \, \Re \int_\p
        n(\ppar{+}k) \, [1{\pm} n(\ppar)] \,
        | {\cal J}_{\ppar\leftarrow\ppar{+}k} |^2 \;
        2 \p_\perp \cdot {\bf f}(\p_\perp;\ppar,\k) \> \theta(\ppar) ,
\label {eq:Wbremw}
\end {equation}
where ${\bf f}(\p_\perp;\ppar,\k)$ is the solution to%
\footnote
    {%
    Rotation invariance about $\k$ implies that
    ${\bf f}(\p_\perp;\ppar,\k)$ must equal $\p_\perp$ times
    a scalar function of $|\p_\perp|$, $\ppar$, and $k$.
    (See Ref.\ \cite{paper2}.)
    But the resulting integral equation (\ref {eq:Wbrem})
    is more compact
    if ${\bf f}(\p_\perp;\ppar,\k)$ is left as a vector function.
    }
\begin {eqnarray}
    2 \p_\perp
    &=&
    i \, \delta E \, {\bf f}(\p_\perp;\ppar,\k)
        + g^2 C_R \int_Q 2\pi\,\delta(q^0{-}\qpar) \,
	\Bigdlangle A^+(Q) \, [A^+(Q)]^* \Bigdrangle \,
\nonumber\\ && \kern 1.7in {} \times
                 \left[{\bf f}(\p_\perp;\ppar,\k)
                       - {\bf f}(\p_\perp{-}\q_\perp;\ppar,\k)\right] .
\label {eq:Wbrem}
\end {eqnarray}
This gives the result for \brem\ production from a single type and
spin (and initial color)
of charged particle in a way that clearly isolates (i) the
dependence on the type of the particle in the splitting factor
$| {\cal J}_{\ppar\leftarrow\ppar{+}k} |^2$, and (ii) the dependence on
the details of the frequency-dependent correlation of the background field
in the correlator $\dlangle A A \drangle$.

If the charge carriers are scalars, one would have
\begin {equation}
   \langle \p | \j_\perp(\k) | \p{+}\k \rangle
   = qe \, \frac{(2\p+\k)_\perp}{(2E_\p)^{1/2} (2E_{\p+\k})^{1/2}}
   \simeq qe \, \frac{\p_\perp}{[\ppar(\ppar+k)]^{1/2}} \,.
\label {eq:sj1}
\end {equation}
The numerator in the middle expression is
the usual photon vertex in relativistic normalization,
while the denominator arises from our use of a non-relativistic normalization
(\ref{eq:QMnorm}) for our states.
Making a similar calculation for fermions, one obtains%
\footnote{
  See, for example, Ref.\ \cite{paper1} for more detail,
  bearing in mind the change to the non-relativistic normalization used here.
}
\begin {equation}
   |{\cal J}_{\ppar\leftarrow\ppar{+}k}|^2 =
   \cases
    {
      \dfrac{1}{4\ppar \, (\ppar{+}k)}_{\strut} ,      & \mbox{scalars;} \cr
      \dfrac{\ppar^2+(\ppar{+}k)^2}{ 8\ppar^2(\ppar{+}k)^2} ,
						      & \mbox{fermions.}
    }
\label {eq:jmatrix}
\end {equation}
This result can also be expressed in terms of the
leading-order Altarelli-Parisi (or DGLAP) splitting
functions for (hard) photon \brem\ from a charged particle $q$,
\begin {equation}
   |{\cal J}_{\ppar\leftarrow\ppar{+}k}|^2 =
   \left.
   \frac{k \, P_{q(\gamma)\leftarrow q}(z)}{8 \ppar^2 (\ppar{+}k)}
   \right|_{z = {\ppar}/(\ppar+k)} ,
\end {equation}
where%
\footnote{
  $P_{q(g)\leftarrow q} = C_R [(1+z^2)/(1-z)_+ + \frac{3}{2} \, \delta(1-z)]$
  is the standard result for gluon \brem\ from a quark.
  The corresponding splitting
  function for photon production is obtained by replacing $C_R$ by 1.
  The treatment of the soft photon
  singularities at $z{=}1$ is not relevant to us, since our photons
  are hard; hence we may ignore the $\delta(1{-}z)$ and the endpoint
  prescription on the $(1{-}z)$ denominator.
}
\begin {equation}
   P_{q(\gamma)\leftarrow q}(z) =
   \cases
   {
      \dfrac{2z}{1-z} ,			& \mbox{scalars;} \cr
      \dfrac{1+z^2}{1-z} ,^{\strut}	& \mbox{fermions,}
   }
\end {equation}
for $z<1$.

% --------------------------------------------------------------------------

\subsection {Relationship to Migdal's equation}

Let us note in passing the relationship of our result to
similar equations written by Migdal \cite{M1,M2} for the case of hard
\brem\ during
electron scattering from the static Coulomb fields of randomly distributed
atoms.  If the background were static, we could write
\begin {equation}
   \Bigdlangle  A^+(Q) \, [A^+(Q)]^*  \Bigdrangle
   =  2\pi \delta(q^0) \> \Bigdlangle  A^+(\q) \, [A^+(\q)]^*  \Bigdrangle ,
\end {equation}
and the collision term in (\ref{eq:SDbrem}) would then become
\begin {equation}
        g^2 C_R \int_\q 2\pi\,\delta(\qpar) \>
        \Bigdlangle A^+(\q) \, [A^+(\q)]^*  \Bigdrangle \,
        [\ff_\nu(\p_\perp;\ppar,\k)-\ff_\nu(\p_\perp{-}\q_\perp;\ppar,\k)]
	\, .
\end {equation}
Migdal has an equivalent expression (Eq.\ (9) of Ref.\ \cite{M2})
but with $\delta(\qpar)$ replaced by
\begin {equation}
   \half \, \delta(E_{\p+\q}-E_\p)
   + \half \, \delta(E_{\p+\k+\q}-E_{\p+\k}) .
\end {equation}
At leading order, however, both of these delta functions are
equivalent to $\delta(\qpar)$, and so our result reduces to Migdal's
in this special case.

Migdal went on to solve his equation in the approximation that the LPM
effect was parametrically large---roughly, that $\delta E \ll
\Gamma_\p$ (or equivalently that the formation time was large compared to
the mean free collision time), in which case it turned out 
he was able to expand in the logarithm of the
ratio,%
\footnote{
   At very high energies, this logarithm is modified in Midgal's analysis
   by the finite nuclear size of the atomic scatterer, which is also
   irrelevant to our application.
}
$\ln(\Gamma_\p / \delta E)$.  In our application, $\delta E$ and $\Gamma_\p$
are of the same parametric order, and such an expansion is not
useful. 

% =========================================================================

\section {Pair annihilation}
\label {sec:pair}

The treatment of pair annihilation is very closely related to that of \brem\
--- a fact which is slightly obscured in the present time-ordered discussion
but is more manifest in our original treatment of Ref.\ \cite{paper1}.
The contribution to the current-current correlator
$W_{\mu\nu}$ due to pair annihilation,
in a form analogous to Eq.\ (\ref{eq:Wbit}) for \brem, is
\begin {equation}
   W_{\mu\nu}^{\rm pair}(K)
   =
   \Biggdlangle
         \int d^4x \> e^{-iKx} \int_{\p\bar\p}
         n(p) \, n(\bar p) \,
           \langle \p \bar\p | j_\mu(0) |\vac\rangle
           \, \langle\vac| j_\nu (x) | \p \bar\p\rangle
   \Biggdrangle \,.
\label {eq:Wbitpair}
\end {equation}
Using the fact that particles propagate independently in our approximation
(the only interactions are with the background field),
this can be written in terms of one-particle evolution amplitudes as
\begin {eqnarray}
   &&
   \int d^4x \> e^{-iKx} \int_{\p\p'\bar\p\bar\p'}
         n(p) \, n(\bar p) \,
   \Biggdlangle
                 \langle \p\bar\p | j_\mu(0) | \vac \rangle
              \, \langle \vac |j_\nu({\bf x}) | \p'\bar\p' \rangle
	      \, \langle \p' \bar\p' | U(x^0,0) | \p \bar\p \rangle
   \Biggdrangle
\nonumber\\ && {} =
   \int d^4x \> e^{-iKx} \int_{\p\p'\bar\p\bar\p'}
         n(p) \, n(\bar p) \>
              \, \langle \p\bar\p | j_\mu(0) | \vac \rangle
              \, \langle \vac |j_\nu({\bf x}) | \p'\bar\p' \rangle
\nonumber\\ && \kern 2in {} \times
              \, \biggdlangle
                   \langle \p' | U(x^0,0) | \p \rangle \,
                   \langle \bar\p' | U(x^0,0) | \bar\p \rangle
              \biggdrangle .
\end {eqnarray}
Translation invariance now gives the momentum conservation constraint
\begin {equation}
   \k = \p + \bar\p = \p' + \bar\p' \, .
\end {equation}
Note that we have ignored processes where the particle and antiparticle
interact directly with each other via gluon exchange, rather than with the
random background field created by other color charge carriers in the plasma.
Such processes are sub-leading in $g$, as are vertex corrections to
\brem.%
\footnote{
   A simple qualitative way to understand this is to consider the
   typical interaction energy times the photon formation time
   (which, among other things, is the time the annihilating pair
   spend within a Debye screening length of each other)
   in the center of mass frame of the almost-collinear annihilating pair.
   If this product is small compared to 1, then the effect of
   direct gluon interactions between the pair is suppressed.
   The formation time is $1/g^2 T$ in the plasma frame, and becomes smaller by
   a factor of $m_\infty/T \sim g$ when boosted to the center-of-mass frame,
   giving $1/g T$.  The characteristic distance between the pair over the
   formation time is $r \sim 1/gT$, giving a characteristic
   potential energy of order
   $g^2/r \sim g^3 T$.  The product is then order $g^2 \ll 1$.  For comparison,
   the interaction time and distance for interaction with a random colored
   particle are
   both order $1/gT$, and the energy-time product is again $g^2$.
   But the pair encounters
   $N \sim 1/g^4$ such random particles during the formation time [number
   density $T^3$ times Debye screening volume $1/(g T)^3$ times formation time
   $1/g^2 T$ divided by time $1/gT$ interacting with each particle].
   These encounters will on average cancel each other in their effects
   except for
   statistical fluctuations, which will enhance our $g^2$ contribution
   per random encounter by
   $\sqrt{N}$ to give a net, unsuppressed effect of $g^2 \sqrt N \sim 1$.
}

\begin{figure}[t]
\includegraphics[scale=0.70]{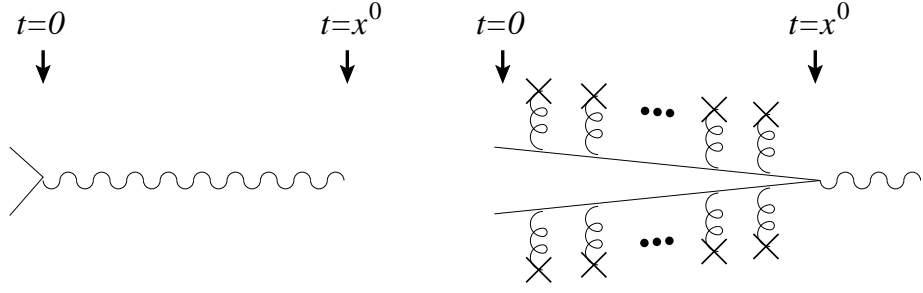}
\caption{\label{fig:pairx}
   Two (time-ordered) diagrams whose interference contributes
   to the pair annihilation rate.
}
\end{figure}

The pair annihilation contribution to the current correlator $W_{\mu\nu}$
represents the interference of the two processes shown in
Fig.\ \ref{fig:pairx}.  The interference of these amplitudes can again
be depicted by a diagram like Fig.\ \ref{fig:jj}, but now the interpretation
is slightly different.  The left photon vertex represents the conjugate of
the first diagram of Fig.\ \ref{fig:pairx}, and everything else represents
the second diagram of Fig.\ \ref{fig:pairx}.
The same time and space ordering considerations
discussed for \brem\ again imply that the only relevant correlations at
leading order are those depicted by Fig.\ \ref{fig:jj2}.
The resulting integral equation is closely related
to the one for \brem:
\begin {equation}
  W_{\mu\nu}^{\rm pair}(K) = 2 \, \Re \int_P
        n(\k{-}\p) \, n(\p) \>
        \langle \vac | j_\mu(\k) | \p, \k{-}\p \rangle^* \>
        \widetilde\fff_\nu(P;K) \, ,
\end {equation}    
with%
\footnote{
  A note on signs: the $i$'s from the two soft gluon interactions cancel the
  minus sign associated with the fact that the color generators for the
  anti-particle are $-T^*$ instead of $T$.
}
\begin {eqnarray}
    \widetilde\fff_\nu(P;K) &=&
    \left( \frac{i}{(k^0{-}p^0) - E_{\k-\p} + i \Gamma_{\k-\p}} \right)^*
    \left( \frac{i}{p^0 - E_\p + i \Gamma_\p} \right)^*
\\ && {} \times
    \left[
	\vphantom{{1\over E_\p}}
        \langle \p, \k{-}\p | j_\nu(\k) | \vac \rangle
        + g^2 C_R \int_Q \biggdlangle
                     [v_{\k-\p}\cdot A(Q)] \,[v_\p\cdot A(-Q)] 
                 \biggdrangle \,
                 \widetilde\fff_\nu(P{-}Q;K)
    \right] .
\nonumber
\end {eqnarray}
The pinching pole approximation is now
\begin {eqnarray}
    \left[ \frac{i}{(k^0{-}p^0) - E_{\k-\p} + i \Gamma_{\k-\p}} \right]^*
       \left[ \frac{i}{p^0 - E_\p + i \Gamma_\p} \right]^*
    &\simeq&
    \frac{ 2\pi i \, \delta(p^0{-}\ppar)
       \> \theta(\ppar) \, \theta(k{-}\ppar)}{
      (E_{\k-\p}+i\Gamma_{\k-\p})+(E_\p+i\Gamma_\p) - k^0 }_{\strut}
\nonumber\\
    &\simeq&
    \frac{ 2\pi \, \delta(p^0{-}\ppar)
	   \> \theta(\ppar) \, \theta(k{-}\ppar)}{
       i \widetilde{\delta E} + (\Gamma_{\k-\p}+\Gamma_\p) } ,
\end {eqnarray}
where
\begin {equation}
   \widetilde{\delta E}
   \equiv - \left[\frac{p_\perp^2+m_\infty^2}{2}\right]
            \left[\frac{k}{\ppar \, (k{-}\ppar)} \right]
   \simeq
   k^0 - E_{\k-\p} - E_\p \,.
\label {eq:dEpair}
\end {equation}
The same simplifications as for \brem\ then reduce these expressions to
\begin {eqnarray}
  \sum_a {\eps^\mu_a}^* \! \eps^\nu_a \> W_{\mu\nu}^{\rm pair}(K) &=&
        2 \, (qe)^2 \, \Re \int_\p \!
        n(k{-}\ppar) \, [1\pm n(\ppar)] \,
        | {\cal J}_{\vac\leftarrow\ppar, k{-}\ppar} |^2 \,
\nonumber\\ && \kern 0.5in {} \times
        2 \p_\perp \cdot
                  \widetilde{\bf f}(\p_\perp;\ppar,\k)
       \> \theta(\ppar) \, \theta(k{-}\ppar) ,
\label {eq:Wpair}
\end {eqnarray}
and
\begin {eqnarray}
    2 \p_\perp
    &=&
    i \, \widetilde{\delta E} \> \widetilde{\bf f}(\p_\perp;\ppar,\k)
        + g^2 C_R \int_Q 2\pi\,\delta(q^0{-}\qpar) \Bigdlangle
                     A^+(Q) \, [A^+(Q)]^* 
                 \Bigdrangle \,
\nonumber\\&& \kern 2in {} \times
	     \left[\widetilde{\bf f}(\p_\perp;\ppar,\k)
		   - \widetilde{\bf f}(\p_\perp{-}\q_\perp;\ppar,\k)\right] ,
\label {eq:Wpaireq}
\end {eqnarray}
where we have defined the small $\p_\perp$ limit
\begin {equation}
   \langle \vac | j_\mu(\k) | \p, \p{-}\k  \rangle
    \simeq 2qe \, \p_\perp \, {\cal J}_{\vac\leftarrow\ppar, k{-}\ppar}
\end {equation}
analogous to Eq.~(\ref{eq:Jlimit}).  These ``joining functions'' are given by
\begin {equation}
   |{\cal J}_{\vac\leftarrow\ppar, k{-}\ppar}|^2 =
   \cases
   {
      \dfrac{1}{4\ppar \,(k{-}\ppar)}_{\strut} ,        & \mbox{scalars}; \cr
      \dfrac{\ppar^2+(k{-}\ppar)^2}{8[\ppar\, (k{-}\ppar)]^2} ,
                                                        & \mbox{fermions}; \cr
   }
\label {eq:jmatrixpair}
\end {equation}
which is the same, up to an overall sign, as taking $\ppar \to -\ppar$ in
the \brem\ formula (\ref{eq:jmatrix}).

The results (\ref{eq:Wpair}) and (\ref {eq:Wpaireq})
can be cast into exactly the same form as
the \brem\ results (\ref{eq:Wbremw}) and (\ref {eq:Wbrem}),
with the only difference being a
factor of $\theta(-\ppar)\,\theta(k+\ppar)$ in Eq.\ (\ref{eq:Wpair}) instead
of $\theta(\ppar)$.  To see this,
everywhere redefine
$\p \to -\p$ in Eq.\ (\ref{eq:Wpair}), including $\ppar \to -\ppar$.
Then define
\begin {equation}
   {\bf f}(\p_\perp,\ppar;\k) \equiv \widetilde{\bf f}(-\p_\perp,-\ppar;\k) .
\end {equation}
The equivalence is completed by the identity
\begin {equation}
   n(-\ppar) = \mp [1 \pm n(\ppar)]
\end {equation}
and the relationship between the squares of current matrix elements
(\ref{eq:jmatrix}) and (\ref{eq:jmatrixpair}) just noted.

% ===========================================================================

\section {Final formulas for photon emission}
\label{sec:final}

To produce a final formula for the leading order contribution to
hard photon production from near collinear \brem\ and pair annihilation,
we just need to combine
the defining relation (\ref{eq:dGamma}) with the result (\ref{eq:Wbrem})
for \brem\ and the result (\ref{eq:Wpair}) for pair annihilation,
transformed as just discussed,
and finally sum over
the colors, spins, and flavors of possible emitters.  The final
result can be neatly packaged in the form
\begin {eqnarray}
   \frac{d\Gamma_\gamma^\LPM}{d^3\k} &=&
    d_R \, \frac{\alphaEM}{4\pi^2 k} \left(\sum_s N_s \, q_s^2\right)
        \int_{-\infty}^{+\infty} \frac{d\ppar}{2\pi}
        \int \frac{d^2\p_\perp}{(2\pi)^2} \;
        n_{\rm f}(\ppar{+}k) \, [1 {-} n_{\rm f}(\ppar)] \,
\nonumber\\ && \kern 2in {} \times
        | {\cal J}_{\ppar\leftarrow\ppar{+}k} |^2 \>
        2 \p_\perp \cdot \Re \, {\bf f}(\p_\perp;\ppar,\k) ,
\label {eq:result1}
\label {eq:result}
\end {eqnarray}
where ${\bf f}(\p_\perp;\ppar,\k)$ satisfies the integral equation
(\ref {eq:Wbrem}), repeated here for convenience,
\begin {eqnarray}
    2 \p_\perp
    &=&
    i \, \delta E \> {\bf f}(\p_\perp;\ppar,\k)
        + g^2 C_R \int_Q 2\pi\,\delta(q^0{-}\qpar) \,
	    \Bigdlangle A^+(Q) \, [A^+(Q)]^* \Bigdrangle \,
\nonumber\\ && \kern 1.7in {} \times
                 \left[{\bf f}(\p_\perp;\ppar,\k)
                       - {\bf f}(\p_\perp{-}\q_\perp;\ppar,\k)\right] ,
\label {eq:result2}
\end {eqnarray}
and $\delta E$ is defined in Eq.~(\ref {eq:dEdef}).
In the result (\ref {eq:result}),
the sum over $s$ is
a sum over the relevant species (flavors) of quarks
({\it i.e.}, $u$, $d$, $s$, $c$, \ldots),
and the current matrix elements are given by the fermionic case of
Eq.~(\ref{eq:jmatrix}).
$N_s=4$ counts the spin and particle/anti-particle states for
each flavor of quark,
and $d_R=3$ is the number of colors of a quark.
As usual,
$n_{\rm f}(\omega) = 1/(e^{\beta\omega}+1)$ is the Fermi distribution
function.
In the result (\ref{eq:result1}),
the contribution to the $\ppar$ integral from $\ppar > 0$
corresponds to \brem\ from quarks,
the $\ppar < -k$ contribution is identical
(after substituting $\ppar \to -k-\ppar$)
and accounts for \brem\ from anti-quarks,
and the $-k < \ppar < 0$ contribution corresponds to pair annihilation.

All dependence on the nature of the scatterers which influence
the photon emission rate is contained in the relevant
$\dlangle AA \drangle$ correlator.
As discussed earlier [{\em c.f.}~Eq.~(\ref{eq:AAcorrelation})],
the background field
correlator is given by $2[n_{\rm b}(q^0)+1] \, \Im \, G_\ret(Q)$,
with the gluon propagator evaluated
in the hard thermal loop approximation for soft momenta $Q \ll T$.
Using standard results for
the hard thermal loop self-energy,
one finds the gauge-invariant%
\footnote
    {%
    The change in the $\langle A^+(Q) A^+(Q)^* \rangle$ correlator
    under an Abelian gauge transformation is proportional to
    $q^+ \equiv q^0 {-} \qpar$, but we are evaluating at $q^+ = 0$.
    Hence this change vanishes for any transformation
    whose Fourier transform is non-singular at $q^+ = 0$
    ({\em e.g.}, any transformation of compact support).
    Because gauge field fluctuations on the $gT$ scale are perturbative,
    the non-Abelian part of a gauge transformation
    changes the correlator by a subleading amount suppressed by $g$.
    }
correlation \cite{paper1}
\begin {equation}
   \Bigdlangle A^+(Q) \, [A^+(Q)]^* \Bigdrangle_{q^0=\qpar}
    	= \frac{\pi \mD^2 T}{2 q}
	\left\{
	    \frac{2}{|q^2 - \PiL(Q)|^2} +
	    \frac{(q_\perp/q)^4}
	{|q^2-(q^0)^2 + \PiT(Q)|^2} \right\}_{q^0=\qpar} ,
\label{eq:Grr}
\end{equation}
when $q^0$ is set equal to $\qpar$
by the delta-function in Eq.~(\ref{eq:result2}).
Here,
$\mD$ is the (lowest-order) color Debye mass,
given by 
$ m_D^2 = (1 + {1\over6} \nf )\, g^2 T^2 $ for QCD with $\nf$ flavors,
and%
\footnote{
   In the convention used here, the hard thermal loop self-energy
   $\Pi^{\mu\nu}(Q) = (Q^2/q^2) \, \PiL(Q) \, {\cal P}^{\mu\nu}_{\rm L}
                    + \PiT(Q) \, {\cal P}^{\mu\nu}_{\rm T}$,
   where ${\cal P}_{\rm L}$ and ${\cal P}_{\rm T}$ are longitudinal
   and transverse projection operators.  Various other authors
   (sometimes including ourselves) instead use the notation
   $\PiL(Q)$ to denote our $- (Q^2/q^2) \, \PiL(Q)$.
}
\begin{eqnarray}
\label{eq:Pi}
\PiL(Q)  &=&
	\mD^2 \left[ -1 + \frac{q^0}{2q} \ln \frac{q+q^0}{q-q^0}
	-i \, \frac{\pi q^0}{2q} \right] ,_{\strut}
\\
\PiT(Q) & = & \mD^2 \, \frac{[q^2-(q^0)^2]}{2q^2} \left[ 
	\frac{(q^0)^2}{q^2-(q^0)^2} + \frac{q^0}{2q} \ln \frac{q+q^0}{q-q^0}
	-i \, \frac{\pi q^0}{2q} \right] .
\end{eqnarray}
Together with Eqs.~(\ref{eq:result1}) and (\ref{eq:result2}),
this reproduces our previous results of Ref.\ \cite{paper1}.

\section{Gluon emission}
\label{sec:gluon}

We now wish to extend of the above analysis to the case of gluon emission.
Other authors who have considered the LPM effect have also discussed
gluon emission \cite{LPM_QCD1,LPM_QCD2,LPM_QCD3,LPM_QCD4,gyulassy&wang}.  
Our generalization is similar to these previous treatments,
but our final result is the first to
give the full correct leading-order answer at high temperature because it
is the first to treat correctly the dynamics of the scatterers.

\begin{figure}[t]
\includegraphics[scale=0.60]{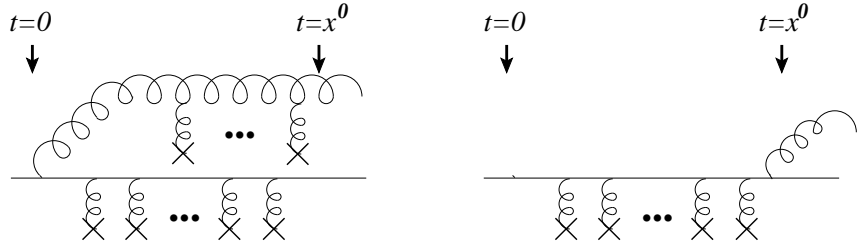}
\caption{\label{fig:bremx2}
   Two diagrams, time ordered from left to right, whose interference
   contributes to the rate of gluon \brem.
   % The first diagram represents gluon emission at time zero,
   % and the second at time $x^0$, and
   % in both cases the diagrams show the evolution between these two times.
   The only difference from the photon emission case of
   Fig.\ \ref{fig:bremx} is that the gluon, as well as the emitter,
   can interact with the soft background field.
}
\end{figure}

\begin{figure}[t]
\includegraphics[scale=0.60]{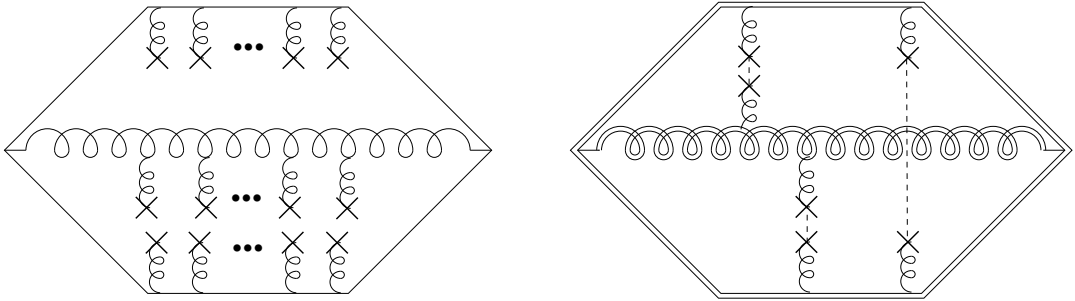}
\bigskip
\caption{\label{fig:jj3}
   A single diagram depicting the interference of the two gluon emission
   amplitudes of Fig.\ \ref{fig:bremx2}.
   The left-hand version depicts the product of amplitudes
   in a given background field;
   the right-hand version shows the result after averaging over
   random background fields.
   Individual background field correlators can connect any 
   two of the three hard lines,
   and double lines represent the resummed propagators
   of Fig.\ \ref{fig:resum}.
   % Each line interacts with the soft
   % background.  When the background is averaged over, then besides 
   % same-line correlations (again, represented by using double
   % lines for the side-bars) there are 3 distinct kinds of 
   % correlations between lines, each illustrated once.
}
\end{figure}

The key difference from photons is that the gluon also carries color.  So,
whereas the photon merely contributed a phase $\exp(i k^0 t)$ to the
time evolution amplitude,
an emitted hard gluon will feel the random colored background field,
just as the emitting particle does.
But provided we consider {\em hard} gluon emission,
the emitted gluon is distinct from the soft background,
and should be treated as another (quasi)particle.
The analog of Fig.\ \ref{fig:bremx} for gluon emission is shown in
Fig.\ \ref{fig:bremx2}.
The analogs of Fig.\ \ref{fig:jj} and Fig.\ \ref{fig:jj2} are shown in 
Fig.\ \ref{fig:jj3}.  Much of the argument for photon emission still
holds; the gluon and quark must be nearly collinear, and consequently the
interactions are ordered in the same way as before. 
The new complications are, first, that there is a color matrix
$T^A_{ab}$ at the hard particle vertex, and second, that there are now
correlations between the soft gauge field felt by the gluon and either emitter
line, not just between the two emitter lines.

Because the interactions are ordered as before, it is still possible to
resum the diagrams by an integral equation similar to Eq.~(\ref{eq:SD1}).
The only difference is that there are now three elements in the collision term,
corresponding to the three kinds of correlations between lines shown in
Fig.\ \ref{fig:jj3}.  The group theoretic coefficients are easily found
with the help of
\begin{equation}
    T_{R}^b \, T_{R}^a \, T_{R}^b
    = 
    \Big( C_{R} - \half \ca \Big) T_{R}^a \, ,
\qquad 
    T_{R}^c \, T_{R}^b \, i f^{abc} = \half \, \ca \, T_{R}^a \, ,
\end{equation}
where $T_{R}^a$ denote representation $R$ color generators.
Since the soft gluon correlators are ordered in time,
this is sufficient to determine the group
factor for the whole diagram;
each line between emitters gives a factor of $(C_{R} - \half\ca)$
and each line from an emitter to the emitted gluon gives $\half \ca$.

The other complication is that, whereas before a cross-rung always changed
$\p$ and left $\k$ the same, now it can
either change $\p$, change $\k$, or change both.
This is a reflection of the fact that the gluon, unlike the photon, can
scatter during the $1/g^2 T$ time scale of the process.
In the photon case, we had a natural fixed direction $\k$ with respect
to which we defined $\p_\perp$ (whose changes we had to keep track of)
and $\ppar$ (whose changes we could ignore).

One inelegant but concrete
possibility
would be to proceed as before but (i) pick by some convention any direction
nearly collinear with $\k$, $\p$ and $\p+\k$
to define the ``parallel'' ($\parallel$)
direction, and (ii) keep track of changes to $\k_\perp$ as well as
$\p_\perp$.  For example, consider picking
the $\parallel$ direction to be that of the momentum of the emitted gluon line
at the right-most corner of Fig.\ \ref{fig:jj3}.
The analog to our results for photon emission would then be%
\footnote{
  $d\Gamma_{\rm g}^\LPM/d^3k$ represents the
  differential
  rate per unit volume of processes that increase by one the number of
  gluons with momentum $\k$.
  One might be tempted to think that
  the integral of this quantity is therefore
  the rate per unit volume of processes
  that increase the total gluon number by one.  This is incorrect
%%  even ignoring the infrared problems and multiple gluon emissions
%%  associated with soft ($\k\ll T$) momenta,
  since pair
  annihilation of hard gluons
  contributes to $d\Gamma_{\rm g}^\LPM/d^3k$ but decreases
  rather than increases the total number of gluons.
  In addition, once one integrates over $\k$, one has to avoid double
  counting final states in the \brem\ process.  We are unaware
  of any simple physical interpretation of the integral of our
  $d\Gamma_{\rm g}^\LPM/d^3k$,
  although this differential rate is closely related to ingredients
  needed in the
  evaluation of various physical quantities, such as leading-order
  transport coefficients \cite {paper3}.
}
\begin {eqnarray}
   \frac{d\Gamma_{\rm g}^\LPM}{d^3\k} &=&
    \frac{\alphas}{4\pi^2k} \sum_s N_s \, d_{s} \, C_{s}
        \int_{-\infty}^{+\infty} \frac{d\ppar}{2\pi} \>
    \Biggl\{
        \int \frac{d^2\p_\perp}{(2\pi)^2} \> \frac{d^2\k_\perp}{(2\pi)^2} \>
        n_{s}(\ppar{+}\kpar) \, [1 \mp n_{s}(\ppar)] \, [1+ n_b(\kpar)]
\nonumber\\ && \kern 1.9in {} \times
        | {\cal J}^{(s)}_{\ppar\leftarrow\ppar{+}\kpar} |^2 \;
        2 \bDelta_{\p\k} \cdot \Re \,
        {\bf f}_s(\p_\perp,\k_\perp;\ppar,\kpar)
    \Biggr\}\Bigg|_{\kpar \to k} ,
\label {eq:inelegant1}
\end {eqnarray}
where we have included a radiation stimulation factor $1{+}n_b(k)$,
defined $\bDelta_{\p\k} \equiv \p_\perp - \ppar \k_\perp/\kpar$ as the
component of $\p$ perpendicular to $\k$ (at leading order), and
where
the accompanying
integral equation is
\begin {eqnarray}
&&
    2 \bDelta_{\p\k} \, (2\pi)^2 \, \delta^{(2)}(\k_\perp)
    =
%\nonumber\\ && \kern 0.2in
    i \, \delta E \> {\bf f}_s(\p_\perp,\k_\perp;\ppar,\kpar)
        + g^2 \int_Q 2\pi\,\delta(q^0{-}\qpar) \>
	    \Bigdlangle A^+(Q) \, [A^+(Q)]^* \Bigdrangle \,
\nonumber\\ && \kern 1.5in {} \times
   \biggl\{
      (C_s - \half \ca)
      \left[{\bf f}_s(\p_\perp,\k_\perp;\ppar,\kpar)
          - {\bf f}_s(\p_\perp{-}\q_\perp,\k_\perp;\ppar,\kpar) \right]
\nonumber\\ && \kern 2.1in {}
      + \half \ca \left[{\bf f}_s(\p_\perp,\k_\perp;\ppar,\kpar)
                      - {\bf f}_s(\p_\perp{+}\q_\perp,\k_\perp{-}\q_\perp;
                                                 \ppar,\kpar)
                                        \right]
\nonumber\\ && \kern 2.1in {}
      + \half \ca \left[{\bf f}_s(\p_\perp,\k_\perp;\ppar,\kpar)
                      - {\bf f}_s(\p_\perp,\k_\perp{+}\q_\perp;\ppar,\kpar)
                  \right]
	\biggr\}.
\end {eqnarray}
The three terms in the integral correspond respectively to gluon exchange
between the top and bottom rails, the top and middle rails, and the bottom
and middle rails of Fig.\ \ref{fig:jj3}.  The momenta flowing from left
to right in the top, middle, and bottom rails of that figure have been
labeled $\p$, $\k$, and $-\p-\k$, respectively.
The energy difference $\delta E$ (for which we have suppressed
the arguments $\p_\perp,\k_\perp,\ppar$ and $\kpar$) is
\begin {equation}
    \delta E
    \equiv
	\frac{m_{\rm g}^2 + k_\perp^2}{2\kpar} + \frac{m_s^2+p_\perp^2}{2\ppar}
	+ \frac{m_s^2+|\p_\perp{+}\k_\perp|^2}{2({-}\ppar{-}\kpar)} 
    \simeq
        E_{\rm g} + E_\p - E_{\p{+}\k} .
\end {equation}
This expression incorporates the fact that in
an equilibrated thermal plasma, the hard gluon dispersion relation
includes an effective mass $m_{\rm g}^2 = \mD^2/2$.

The preceding is a valid but inelegant approach because it does not
exploit the rotational invariance of the problem.
Other than $|{\bf f}_s|$, the elements of the integrand in
the rate formula (\ref{eq:inelegant1}) are invariant under small angle
($\theta \sim g$) rotations, which do not change $\ppar$ and $\kpar$ at
leading order, but do change $\p_\perp$ and $\k_\perp$.
So one could split the $d^2p_\perp \> d^2k_\perp$ into one integral
with respect to the relative momentum $\Delta_{\p\q}$, which appears
explicitly in (\ref{eq:inelegant1}), and another integral over small
rotations.  The integration over small rotations could then be
absorbed into
${\bf f}_{\rm s}$, defining a new function that depends only
on $\Delta_{\p\q}$.
It turns out that the most symmetric procedure does not use $\Delta_{\p\q}$
but rather the quantity $\D \equiv \kpar\Delta_{\p\q}$,
which can be written as
\begin {equation}
   \D = \kpar \, \p_\perp - \ppar \, \k_\perp .
\end {equation}
This may also be written as
$\D = (\k\times \p) \times \epar$,
where $\epar$ is the unit vector in the $\parallel$ direction.
The corresponding equations that only track changes in $\D$ are
\begin {mathletters}
\label {eq:gfinal}
\begin {eqnarray}
    \frac{d\Gamma_{\rm g}^\LPM}{d^3\k} &=&
    \frac{\alphas}{4\pi^2 k^2} \sum_s N_s \, d_{s} \, C_{s}
        \int_{-\infty}^{+\infty} \frac{dp}{2\pi}
        \int \frac{d^2 \D}{(2\pi)^2} \>
        n_{s}(p{+}k) \, [1 \mp n_{s}(p)] \, [1+ n_b(k)]
\nonumber\\ && \kern 2in {} \times
	\frac{1}{k^3} \,
        | {\cal J}^{(s)}_{p\leftarrow p{+}k} |^2 \;
        2 \D \cdot \Re \, {\bf F}_s(\D;p,k) \,,
\label {eq:result3}
\label {eq:resultt}
\end {eqnarray}
where ${\bf F}_s(\D;p,k)$ is the solution to the integral equation
\begin {eqnarray}
    2 \D
    &=&
    i \, \delta E(\h;p,k) \, {\bf F}_s(\D;p,k)
        + g^2 \int_Q 2\pi\,\delta(q^0{-}\qpar) \>
	    \Bigdlangle A^+(Q) \, [A^+(Q)]^* \Bigdrangle \,
\nonumber\\ && \kern 1in {} \times
	\biggl\{
	     (C_s - \half \ca) \left[{\bf F}_s(\D;p,k)
                       - {\bf F}_s(\D{-}k\,\q_\perp ;p,k) \right]
\nonumber\\ && \kern 1.6in {}
	     + \half \ca \left[{\bf F}_s(\D;p,k)
                       - {\bf F}_s(\D{+}(k{+}p)\q_\perp;p,k) \right]
\nonumber\\ && \kern 1.6in {}
	     + \half \ca \left[{\bf F}_s(\D;p,k)
                       - {\bf F}_s(\D{-}p\,\q_\perp;p,k) \right]
	\biggr\},
\label {eq:foo}
\end {eqnarray}
with
\begin {equation}
    \delta E(\h;p,k)
    =
	\frac{m_{\rm g}^2}{2k} + \frac{m_s^2}{2p}
	+ \frac{m_s^2}{2({-}p{-}k)} 
	- \frac{\h^2}{2p \, k\, ({-}p{-}k)} \, .
\label {eq:result4}
\end {equation}
\end {mathletters}%
To simplify the appearance of these results, and the equations below,
we have dropped the $\parallel$ subscripts on $\ppar$ and $\kpar$.
It should be emphasized that evaluating these expressions does not,
in fact, require selecting any particular convention for defining
the $\parallel$ direction.
Also, note that rotation invariance implies that
${\bf F}_s(\D;p,k)$ must equal $\D$ times
a scalar function of $|\D|$, $\ppar$, and $k$,
but the integral equation (\ref {eq:foo})
is more compact if ${\bf F}_s(\D;p,k)$ is left as a vector function.

In the result (\ref {eq:result3}),
the sum over species $s$ now includes gluons, which have
$N_s=2$, $d_s=d_{\rm A}$, and $C_s = \ca$
(with $d_{\rm A}{=}8$ and $\ca{=}3$ for QCD).
The splitting function for gluons is
\begin{equation}
P_{gg \leftarrow g}(z) = \frac{1+z^4+(1{-}z)^4}{z\, (1{-}z)}
\end{equation}
for $z < 1$, so that
\begin{equation}
| {\cal J}^{(g)}_{p\leftarrow p{+}k} |^2 = \frac{p^4 + k^4 + 
	(p{+}k)^4}{8 p^3 \, (p{+}k)^3} \, .
	%\qquad \mbox{gluons} \, .
\end{equation}
This expression, when multiplied by $k^{-3}$,
is symmetric under interchanges of the three gluon momenta
$p$, $k$, and ${-}p{-}k$.
%%The fact that the inverse fourth power of $k$ appears
%%in Eq.~(\ref{eq:result3}) simply reflects
%%our choice to write $d\Gamma_{\rm g}/d^3 \k$ rather
%%than the Lorentz invariant measure $k \, d\Gamma_{\rm g}/d^3 \k$.

Further symmetry of our result (whether for quark or gluon emitters)
can be made
manifest by rewriting group factors so that the expression in
curly braces in the integral equation (\ref{eq:foo}) is
\begin {eqnarray}
	\biggl\{
	     \half (C_{R_3} + C_{R_1} - C_{R_2}) && \left[{\bf F}_s(\D;p,k)
                       - {\bf F}_s(\D{-}p_2 \, \q_\perp;p,k) \right] + {}
\nonumber\\
	     \half (C_{R_1} + C_{R_2} - C_{R_3}) && \left[{\bf F}_s(\D;p,k)
                       - {\bf F}_s(\D{-}p_3 \, \q_\perp;p,k) \right] + {}
\nonumber\\
	     \half (C_{R_2} + C_{R_3} - C_{R_1}) && \left[{\bf F}_s(\D;p,k)
                       - {\bf F}_s(\D{-}p_1 \, \q_\perp;p,k) \right]
	\biggr\},
\label{eq:prettygroup}
\end {eqnarray}
where $(p_1,p_2,p_3)$ denotes the three momenta $(p,k,{-}p{-}k)$ and
$(C_{R_1},C_{R_2},C_{R_3})$ denote the quadratic Casimirs
$(C_s,C_{\rm A},C_s)$ of the corresponding color
representations.%
\footnote{
   There is a general argument that gives the structure of the group
   factors shown in Eq.~(\ref{eq:prettygroup}) even if $R_1$, $R_2$, and
   $R_3$ were any three color representations.  Color conservation at
   either end of Fig.\ \ref{fig:jj3} means that these three
   representations have to be combined into a color singlet.  Acting
   on that color singlet state with color generators
   $T^a_{R_1} + T^a_{R_2} + T^a_{R_3}$ should
   give zero.  So acting with $T^a_{R_1} + T^a_{R_2}$ has the same effect
   as acting with $- T^a_{R_3}$.  Squaring both sides and summing over
   the basis $a$ of the algebra, one then obtains that
   $T_1^a T_2^a$
   (the factor associated
   with exchanging a gluon between particles 1 and 2),
   acting on the color singlet,
   is equivalent to
   multiplication by $\half(C_3-C_1-C_2)$.  The other group factors in
   (\ref{eq:prettygroup}) follow by permutation.
}

The previous photon emission result can be extracted from the final
gluon emission
expressions (\ref{eq:gfinal}) by setting $\ca{=}0$ inside the integral equation
(which corresponds to making the emitted particle colorless
rather than in the adjoint representation), 
replacing $C_s$ in Eq.~(\ref{eq:result3}) with $q_s^2$,
and using the photon,
rather than gluon, thermal mass in the expression (\ref {eq:result4})
for $\delta E$.
(The resulting photon thermal mass contribution to $\delta E$ is
irrelevant for hard photons \cite {paper1}.)

Finally, we make a last note about the cancellation of infrared divergences
in our result for gluon production.  The fact that the
%% functions ${\bf F}_{\rm s}$
%% far-infrared-canceling combinations
%% ${\bf F}_s(\D;p,k)-{\bf F}_s(\D-(\cdots)\q_\perp;p,k)$ in (\ref{eq:foo})
%% or (\ref{eq:prettygroup})
%
collision term in (\ref{eq:foo}) only involves differences of the form
${\bf F}_s(\D;p,k)-{\bf F}_s(\D-p_i\q_\perp;p,k)$,
which cancel in the far-infrared ($\q_\perp\to 0$) limit,
results from including all three ways of connecting the three
rails in Fig.\ \ref{fig:jj3} together with resumming self-energies
(as in Fig.\ \ref{fig:resum}) on all three rails.
In contrast, if we had truncated the hard gluon line and
evaluated the earlier diagrams of Fig.\ \ref{fig:jj2}
with external color currents,
these cancellations would have been absent and our
result would have had far-infrared divergences signalling sensitivity
to ultra-soft momentum scales.
The sensitivity of such truncated diagrams to ultra-soft physics
is relevant to the evaluation of quantities such as
color conductivity,%
\footnote{
   %% This is a ridiculous amount of referencing for such a small comment,
   %% but I couldn't figure out where to stop without offending someone.
   For a discussion of color conductivity in terms of ladder diagrams, see
   Ref.\ \cite{ccdiagrams}.
   For an equivalent discussion in terms of kinetic theory, see
   Ref.\ \cite{ccus} and the pathbreaking work of Ref.\ \cite{bodeker}.
   See also Refs.\ \cite{cc1,cc2} and the original,
   early work of Ref.\ \cite{selikhov}.
}
but not to the problem of the hard gluon emission rate at leading order.

% =========================================================================

\begin{acknowledgments}

This work was supported, in part, by the U.S. Department
of Energy under Grant Nos.~DE-FG03-96ER40956
and DE-FG02-97ER41027.
We thank Francois Gelis and Patrick Aurenche for helpful conservations.

\end{acknowledgments}

% =========================================================================

\end{document}